\documentclass[11pt,aps,amsmath,amssymb,nofootinbib,notitlepage,longbibliography]{revtex4-1}
\usepackage{amsmath,amssymb}
\usepackage{slashed}
\usepackage{graphicx}
\usepackage{bm}
\usepackage[top=1.0in,bottom=1.0in,left=1.0in,right=1.0in]{geometry}
\usepackage[colorlinks,linkcolor=blue,citecolor=blue]{hyperref}
\usepackage{subcaption}
\newcommand{\be}{\begin{equation}}
\newcommand{\ee}{\end{equation}}
\newcommand{\bea}{\begin{eqnarray}}
\newcommand{\eea}{\end{eqnarray}}
\newcommand{\ba}{\begin{eqnarray}}
\newcommand{\ea}{\end{eqnarray}}

\begin{document}

\title{Mueller's dipole wave function in QCD: \\ emergent KNO scaling in the double logarithm limit}

\author{Yizhuang Liu}
\email{yizhuang.liu@uj.edu.pl}
\affiliation{Institute of Theoretical Physics, Jagiellonian University, 30-348 Krak\'{o}w, Poland}

\author{Maciej A. Nowak}
\email{maciej.a.nowak@uj.edu.pl}
\affiliation{Institute
of Theoretical Physics and Mark Kac Center for Complex Systems Research,
Jagiellonian University, 30-348 Krak\'{o}w, Poland}

\author{Ismail Zahed}
\email{ismail.zahed@stonybrook.edu}
\affiliation{Center for Nuclear Theory, Department of Physics and Astronomy, Stony Brook University, Stony Brook, New York 11794--3800, USA}

\begin{abstract}
We analyze Mueller's QCD dipole wave function evolution in the double logarithm approximation (DLA). Using complex analytical methods, we show that the distribution of dipole in the wave function (gluon multiplicity distribution) asymptotically satisfies the Koba-Nielsen-Olesen (KNO) scaling, with a non-trivial scaling function $f(z)$ with $z=\frac n{\bar n}$.  The scaling function decays exponentially as $2(2.55)^2ze^{-\frac{z}{0.3917}}$ at large $z$, while its growth is log-normal as $e^{-\frac{1}{2}\ln^2 z}$  for small-$z$.  A detailed analysis of the Fourier-Laplace transform of $f(z)$, allows for performing  the inverse Fourier transform,  and access the non-asymptotic bulk-region around the peak. The bulk and asymptotic results
 are shown to be in good agreement with the measured hadronic multiplicities in DIS, as reported by the H1 collaboration at HERA in the region of large $Q^2$.   A numerical tabulation  of $f(z)$ is included. Remarkably, the same scaling function is found to emerge in the resummation of double logarithms in the evolution of jets. Using the generating function approach, we show why this is the case.  The absence of KNO scaling in non-critical and super-renormalizable theories is briefly discussed.
 We also discuss the universal character of the entanglement entropy in the KNO scaling limit,  and its measurement using the  emitted multiplicities  in DIS and
 $e^+e^-$ annihilation.
 \end{abstract}

\maketitle

\section{Introduction}
In a broad and general sense, the high energy limit of QCD is non-trivial. In the simplest case of  $e^+e^-$ annihilation,  asymptotic freedom
is sufficient  to guarantee a controlled $\ln^{-1} \frac{Q^2}{\Lambda_{\rm QCD}^2}$ expansion in leading power~\cite{Politzer:1973fx,Gross:1973id}. However, there is  a number of  examples where the running coupling constant is not the only source of large logarithms. Remarkably, a complete understanding of  the high-energy limit in these situations,  is still challenging, 50 years after the discovery of asymptotic freedom.

One important example which is relatively easy to understand is the Bjorken limit~\cite{Bjorken:1968dy}, of processes such as DIS~\cite{Bloom:1969kc} and DVCS~\cite{Ji:1996nm} at moderate parton $x$, where it is commonly accepted~\cite{Sterman:1993hfp,Collins:2011zzd} that in leading power,  the structure functions (at least, its moments) can be factorized into IR-sensitive matrix elements times hard coefficients in a way that can be cast into a controlled large $Q^2$ expansion of the form $\alpha(Q)^{\frac{\gamma_1}{\beta_0}}(1+{\cal O}[\alpha(Q)])c_n$ with the help of renormalization group equation (RGE). A more challenging situation is the Regge limit or small-$x$ limit, where there are rapidity logarithms of the type  $\ln \frac{1}{x}$, that cannot be systematically controlled by the RGE formalism.  Nevertheless, if one is only interested in resumming the ``leading logarithms'' $\alpha_s^n\ln^n \frac{1}{x}$ in pQCD (assuming all the $k_\perp$'s are large), then major progresses have been made. In particular, the leading rapidity logarithms in the light-front wave functions (LFWFs) of a color-dipole (Mueller$^\prime$s dipole),  can be effectively extracted using  Mueller$^\prime$s evolution equation~\cite{Mueller:1993rr,Mueller:1994gb}. This is essentially a cascade formed by consecutive small-$x$ gluon emissions ordered in rapidity. From Mueller$^\prime$s dipole, many other evolution equations such as the linearized BFKL equation~\footnote{Notice that the BFKL equation is more universal than Mueller's dipole. Indeed, the BFKL spectrum is related to analytical continuation in anomalous dimensions of twist-two operators, therefore generalizable to ${\cal N}=4$ SUSY CFT at a generic gauge coupling.} for the gluon density~\cite{Kuraev:1977fs,Balitsky:1978ic,Lipatov:1996ts}, or its BK variant~\cite{Balitsky:1995ub,Kovchegov:1999yj} for the cross-section can be derived under further assumptions.

Nevertheless, whether the small-$x$ pQCD evolutions present a correct and self-consistent description of the Regge limit is still not clear. It is well-known that the BFKL evolution mixes different ``twists'' and tends to ``diffuse'' into soft regions, which is made worse  by the ``IR renormalon'' caused by running inside massless integrals. It is also well-known,  that the derivation of the famous Froissart's bound relies crucially on the existence of an exponential decay at large impact parameter, a feature that is essentially non-perturbative. The popular assumption is that there will be an emergent ``saturation scale'' $Q_s$,   that is hard enough to justify pQCD at large rapidity or small $x$,  due to
ever increasing  gluon density. As a result, the confinement problem can be avoided. The ``color-glass condensate'' (CGC) approach~\cite{McLerran:1993ni,McLerran:1993ka,Iancu:2002xk,Gelis:2010nm} to gluon saturation is based on this assumption. The string-gauge duality provides another way to look at the Regge limit at strong coupling (large impact parameter or small $Q^2$). It has been proposed that the wee dipoles are string bits~\cite{Susskind:1993ki,Susskind:1993aa,Thorn:1994sw}, and their evolution
is best captured by a dual string~\cite{Rho:1999jm,Janik:2000aj,Polchinski:2002jw,Brower:2006ea,Basar:2012jb}. In~\cite{Basar:2012jb,Stoffers:2012zw}, a string-based holographic Reggeization formalism in which forward dipole-dipole scattering is realized as the exchange minimal surfaces in appropriate geometries has been proposed and agrees qualitatively with the Mueller$^\prime$s approach~\cite{Mueller:1993rr,Mueller:1994gb} in the conformal limit. In the large impact parameter limit, due to the presence of a natural string tension, the approach fulfills the Froissart's bound in a non-perturbative manner.

Among all the features of the small-$x$ evolution, the growth of the ``parton number'' with rapidity plays a key role.  In the Mueller$^\prime$s dipole picture, the LFWF of the projectile dipole tends to split into more and more dipoles ordered strongly in rapidity, before interacting with the target. The large number of color charges present in the wave function, should be responsible for the large number of observed particle multiplicity~\cite{H1:2020zpd}. The distribution of the virtual dipoles in the Mueller$^\prime$s wave function in the ``diffusion limit'' has long been believed to be similar to the simplified $0+1$D reduction~\cite{Mueller:1994gb,Gotsman:2020bjc,Levin:2021sbe}, which satisfies the famous KNO scaling~\cite{Polyakov:1970lyy,Koba:1972ng}, with a ``geometric'' scaling function $e^{-z}$. However, this scaling function differs significantly from the observed scaling function for ep data~\cite{H1:2020zpd}. On the other hand, the dipole evolution has another limit, which actually forms the common region with the DGLAP evolution~\cite{Kovchegov:2012mbw}, the double logarithm approximation (DLA) limit. In this limit, the growth of the gluon density is much slower, however, as we will show in this paper, the distribution of dipoles in the wave function displays a non-trivial scaling function, in agreement with the reported hadronic multiplicities  by the H1 collaboration at HERA~\cite{H1:2020zpd}.

It has been suggested~\cite{Stoffers:2012mn,Kharzeev:2017qzs}, that underlining the large number of observed particle multiplicity, is the onset of a quantum or entanglement entropy. In~\cite{Kharzeev:2017qzs,Levin:2021sbe}, the authors argued that the Mueller$^\prime$s wavefunction is strongly entangled in longitudinal momentum space on the light front. This
entanglement is distinct from the spatial entanglement, usually encoded in the ground state wave function in the rest frame~\cite{Srednicki:1993im,Calabrese:2004eu,Casini:2005rm,Hastings:2007iok,Calabrese:2009qy}.  In the large rapidity $y$ limit, this entanglement is
universally captured by an entropy $S= {\rm ln} \bar n\sim y$, where the mean multiplicity $\bar n\sim e^{\# y}$ grows exponentially,
as noted in  the dual string~\cite{Stoffers:2012mn}, and Mueller$^\prime$s cascade~\cite{Kharzeev:2017qzs}.
These observations, have attracted a number of recent studies both theoretically~\cite{Shuryak:2017phz,Kovner:2018rbf,Beane:2018oxh,Cloet:2019wre,Armesto:2019mna,Gotsman:2020bjc,Kharzeev:2021nzh,Levin:2021sbe,Liu:2022ohy,Dumitru:2022tud, Liu:2022hto,Ehlers:2022oal},
and empirically~\cite{Kharzeev:2021yyf,Hentschinski:2021aux,Hentschinski:2022rsa}. For completeness, we note that a classical thermodynamical  entropy  using the production of gluons at high
energy, was initially explored in~\cite{Kutak:2011rb}.

Recently in~\cite{Liu:2022ohy,Liu:2022hto}, we have formalized the {\it rapidity space entanglement} between fast and slow degree of freedom in the Mueller$^\prime$s dipole, and derived a Balitsky-Kovchegov (BK) type equation for the associated reduced density matrix in the large $N_c$ limit. We have shown explicitly that for both the $0+1$D reduction and  the $1+2$D QCD, the eigenvalues of the {\it reduced density matrix} indeed coincide with the dipole distribution. As a result, we have shown that the multiplicity entropy is of quantum nature. In particular, in the $0+1$D reduction the dipole multiplicities follow a simple exponential distribution as in~\cite{Mueller:1994gb}, while in the non-conformal QCD in $1+2$ dimensions,  the mean  dipole multiplicities were found to follow a Poisson distribution $p_n=e^{-\bar n} {\bar n}^n/n!$,  with a linear growth of mean multiplicity with rapidity. The quantum entanglement entropy at large rapidity,  asymptotes $S=\frac 12 {\rm ln}\bar n\sim  \frac 12 {\rm ln}y$,  which is much smaller  than in $0+1$D reduction.
The cascade of dipoles in $1+2$D dimensions is ``quenched''  kinematically by transverse integrals, and provides a simple mechanism for saturation.

The solution of the evolved BK type equations for the density matrix, and the underlying  multiplicity of the emitting dipoles for QCD in $1+3$ dimensions,
is not known except in the diffusion limit~\cite{Mueller:1994gb,Gotsman:2020bjc,Levin:2021sbe}. Needless to say, that these  dipole or gluon multiplicities, when released in a prompt ep or pp collision at high energy,
are of relevance to the measured hadronic multiplicities.
The purpose of this paper is to address this open problem partially, by solving directly Mueller$^\prime$s evolution in the double logarithm approximation (DLA), which re-sums the large ${\rm ln}\frac 1x$
and large ${\rm ln} \frac{Q^2}{\Lambda_{\rm QCD}^2}$ simultaneously. This allows for an explicit
derivation of the wee dipole distributions, which will turn out to be in  good agreement with the currently available H1 data from DIS scattering at HERA~\cite{H1:2020zpd}. It also provides for an
estimation of the entanglement entropy for DIS scattering in QCD, a  measure of gluon decoherence and possibly saturation. We should emphasize that the DLA limit lies in the common region of DGLAP evolution, and Mueller's evolution. Hence, all the results in this paper can also be derived using the wave function evolution~\cite{Burkardt:2002uc,Kovchegov:2012mbw},  underlining the DGLAP equation as well.

It  is worth noting that the scaling function we established for Mueller$^\prime$ wave function evolution, appears in  the context of jet evolution~\cite{Dokshitzer:1982ia,Dokshitzer:1991wu}. In fact, this is not a coincidence as we will detail below,   and follows  from the BMS-BK correspondence, that  maps the IR divergences in $e^+e^-$ annihilation to the rapidity divergences in the dipole's wave function~\cite{Banfi:2002hw,Weigert:2003mm,Marchesini:2003nh,Hatta:2013iba,Caron-Huot:2015bja}. In leading order, this follows from the fact that all the leading IR logarithms are generated by a Markov process of consecutive emission of soft gluons, strongly ordered in energy into the asymptotic final state cuts, in a way very similar to the underlining branching process of the Mueller's dipole. At  leading order, the two evolutions are related by a conformal transformation~\cite{Cornalba:2007fs,Vladimirov:2016dll}, that maps light-like dipoles to Wilson-line cusps, and virtual soft gluons in LFWF to real gluons in asymptotic states. The BMS evolution has a natural double logarithm limit corresponding to the Sudakov double-logarithm in the virtual part (form-factor), that can be obtained by imposing angular ordering on the emitted soft gluons.  Through a conformal transformation, the angular-ordering  maps to the dipole size ordering. As a result,  the scaling function of the two DL limits are identical. These observations allow us to extend the concept of entanglement to jet evolution as well.

The organization of the paper is as follows. In section~\ref{SEC_DLA} we simplify the Mueller$^\prime$s evolution equation for the generating functional in the DLA limit. We show that the resulting generating function in DLA limit satisfies a second order non-linear differential equation, similar but not equivalent to the Painlev\'{e}-III equation.  For large mean multiplicities $\bar n$, this allows the determination of all the leading consecutive moments of the dipole distribution, through a second order recursive hierarchy. We study the behavior of the moment sequence,  and show that the underlying multiplicity distribution obeys Koba-Nielsen-Olesen (KNO) scaling~\cite{Polyakov:1970lyy,Koba:1972ng} in the form $p_n=\frac{1}{\bar n}f\left(\frac{n}{\bar n}\right)$, with scaling variable $z=n/\bar{n}$ in large $\bar n$ limit and the scaling function $f(z)$. In section~\ref{SEC_KNO}, we show that the complex analytic Fourier-Laplace transform $Z(t)$ of the KNO scaling function $f(z)$ can be determined by analytically continuing  from a simple integral representation, from which the $f(z)$ can be obtained by Fourier-inversion. In particular, for large and small $z$, the asymptotic forms of $f(z)$ can be determined exactly,  and for general $z$ numerically. In section~\ref{SEC_H1} we compare our DLA scaling function $f(z)$ to the empirical charged multiplicity scaling function $\Psi(z)$ extracted from ep data at HERA~\cite{H1:2020zpd},  and find good agreement.  In section~\ref{SEC_BMS}, we briefly review  the leading order BMS evolution equation, using the generating functional formalism of~\cite{Mueller:1993rr}. The relation to the dipole evolution is made manifest. In section~\ref{SEC_ENT} we show that the ensuing multiplicities for both the wavefunction and jet evolutions are quantum entangled. The entanglement entropy asymptotes $S={\rm ln} \bar n$ in the DLA.
The logarithmic growth of the entanglement entropy with $\bar n$,  is generic for all hadronic multiplicities in QCD with  KNO scaling. Our conclusions are in~\ref{SEC_CON}. In the Appendix, we briefly discuss the multiplicity distributions of super-renormalizable theories, and their lack of KNO scaling.

\section{Mueller's dipole wave function and its DLA limit}
\label{SEC_DLA}
In this section we consider  Mueller$^\prime$s dipole evolution equation in the DLA limit. We briefly recall that in~\cite{Mueller:1993rr,Mueller:1994gb} using
the planar limit, it was shown that the consecutive emission of gluons with smaller and smaller $x$, into the light-front wave function (LFWF) of a valence quark-anti-quark pair (the Mueller's dipole), leads to a closed equation for the generating functional of the squared norms of the LFWF
\begin{align}\label{eq:evoZ}
&{\cal Z}(b_{10},\frac{x_0}{x_{\rm min}},\lambda)=S(b_{10},\frac{x_0}{x_{\rm min}})\nonumber \\
&+\lambda \frac{\alpha_s N_c}{2\pi^2} \int_{x_{\rm min}}^{x_0} \frac{dx_1}{x_1} S(b_{10},\frac{x_0}{x_1}) \int db_2^2\frac{b_{10}^2}{b_{12}^2b_{20}^2}{\cal Z}(b_{12},\frac{x_1}{x_{\rm min}},\lambda){\cal Z}(b_{20},\frac{x_1}{x_{\rm min}},\lambda) \ ,
\end{align}
with the Sudakov or  ``soft-factor'' for ``virtual'' emissions\footnote{The use of the words ``virtual'' and ``real'' in the context of wave functions,  is not very precise. A more precise definition would be   ``disconnected'' and ``connected'' contributions.  This said, note that this contribution is the square of the standard transverse momentum dependent (TMD) soft-factor, one factor from the wave function and the other factor from the conjugate wave function.} as
\begin{align}
\label{SUDAKHOV}
S\bigg(b_{10},\frac{x_0}{x_1}\bigg)=\exp \bigg[-\frac{\alpha_s N_c}{\pi}\ln b_{10}^2\mu^2 \ln \frac{x_0}{x_1}\bigg] \ .
\end{align}
$Z$   generates the probability of finding $n+1$ dipole inside the LFWF of the $Q\bar Q$ pair
\begin{align}
\label{TAYLOR}
{\cal Z}(b,y,\lambda)=\sum_{n=0}^{\infty}\lambda^n p_n(b,y)  \ .
\end{align}
Unitarity requires $Z=1$ for $\lambda=1$, which is manifest in (\ref{TAYLOR}). The {\it factorial moments}~\footnote{Sometimes they are also called the ``disconnected moments'' or ``multiplicity correlators''.} of
the distribution  $p$ follows as
\begin{align}\label{eq:expect}
\frac{d^k}{d\lambda^k}{\cal Z}(b,y,\lambda)|_{\lambda=1}=\sum_{n=0}^{\infty}n(n-1)..(n-k+1)p_n \equiv \langle n(n-1)...(n-k+1) \rangle \ .
\end{align}
The knowledge of ${\cal Z}$ provides a  detailed understanding of the LFWF of the $Q\bar Q$ pair, in the small-$x$ sector.

\subsection{The double logarithm limit}
The double logarithm limit (DLA) corresponds to the situation where $b_2$ is very close to either $b_0$ or $b_1$, the locations of the mother dipole and so on with $b_3$......
As a result, the emitted dipoles carry smaller and smaller sizes. In this limit, if one introduce the scale parameter $\alpha=\ln \frac{b_{10}^2}{b^2}=\ln \frac{Q^2}{Q_0^2}$ where $b$ is the size of the emitted dipole, which is  identified as inverse of $Q$, then the evolution equation simplifies to
\begin{align}\label{eq:evoZDLA}
&{\cal Z}(y,\alpha,\lambda)=\exp \bigg[-\frac{\alpha_s N_c}{\pi}\alpha y\bigg]\nonumber \\
&+\frac{\alpha_s N_c\lambda}{\pi}\int_{0}^{\alpha}d\alpha'\int_{0}^y dy'\exp\bigg[-\frac{\alpha_sN_c}{\pi}\alpha (y-y')\bigg]{\cal Z}(\alpha',y',\lambda){\cal Z}(\alpha,y',\lambda) \ .
\end{align}
It is easy to check that for $\lambda=1$ one has the trivial solution ${\cal Z}=1$. By taking derivative with respect to $y$, one obtains the exact equation in DLA
\begin{align}
\frac{\partial {\cal Z}(\alpha,y)}{\partial y}=\frac{\alpha_s N_c}{\pi}{\cal Z}(\alpha,y)\bigg(-\alpha+\lambda\int_{0}^\alpha d\alpha' {\cal Z}(\alpha',y)\bigg) \ .
\end{align}
On the other hand, for the mean  $\bar n=\frac{d{\cal Z}}{d\lambda}|_{\lambda=1}$,
\begin{align}
\frac{\partial \bar n}{\partial y}=\frac{\alpha_sN_c}{\pi}\bigg(\alpha+\int_{0}^{\alpha} \bar n(\alpha',y)d\alpha'\bigg) \ , \\
\frac{\partial^2 \bar n}{\partial y \partial \alpha}=\frac{\alpha_sN_c}{\pi}(\bar n+1) \sim \frac{\alpha_sN_c}{\pi} \bar n \ .
\end{align}
The second equation is nothing but the DGLAP evolution equation in the DLA limit for the gluon density, allowing the identification $\bar n=xG(x,Q^2)$.

Note that for a running gauge coupling $\alpha_s$,
\begin{align}
\label{AG}
\alpha_s(Q)=\frac{1}{\beta_0 \ln \frac{Q^2}{\Lambda_{\rm QCD}^2}} \qquad{\rm and}\qquad \gamma=\ln \frac{\ln \frac{Q^2}{\Lambda_{\rm QCD}^2}}{\ln \frac{Q_0^2}{\Lambda_{\rm QCD}^2}} \ ,
\end{align}
where $\beta_0=\frac{1}{4\pi}(\frac{11N_c}{3}-\frac{2N_f}{3})$,  the above equations still hold, with
\begin{align}
{\cal Z}(y,\gamma)=\exp\bigg[-\frac{N_c}{\pi \beta_0}y\gamma \bigg]+\lambda \frac{N_c}{\pi \beta_0}\int_{0}^{\gamma}d\gamma'\int_0^ydy' \exp \bigg[-\frac{N_c}{\pi \beta_0}(y-y')\gamma\bigg]{\cal Z}(y',\gamma'){\cal Z}(y',\gamma) \ ,
\end{align}
which is identical to Eq.~(\ref{eq:evoZDLA}) with the identification $\alpha_s \rightarrow \frac{1}{\beta_0}$ and $\alpha \rightarrow \gamma$. Below we still use the notation in Eq.~(\ref{eq:evoZDLA}) with this understanding.

To investigate the multiplicity distribution in the DLA, we note that  the generating function ${\cal Z}$ is only a function of $\rho$,
\begin{align}
&\rho=\frac{\alpha_sN_c}{\pi}y \ln \frac{Q^2}{Q_0^2} \ ,\  {\rm no  \ running} \ , \\
&\rho=\frac{N_c}{\pi \beta_0}y \ln \frac{\ln \frac{Q^2}{\Lambda_{\rm QCD}^2}}{\ln \frac{Q_0^2}{\Lambda_{\rm QCD}^2}}\ ,\  {\rm with \ running}  \ ,
\end{align}
which amounts to
\begin{align}
{\cal Z}(\rho)=e^{-\rho}+\lambda \rho \int_{0}^1 dt_1\int_{0}^1 dt_2 e^{-\rho(1-t_2)}{\cal Z}(\rho t_1t_2){\cal Z}(\rho t_2) \ ,
\end{align}
in both cases.
In particular, the equation for the averaged number of soft gluons $\bar n=xG(x,Q)$ becomes
\begin{align}
\rho \frac{d^2 \bar n}{d\rho^2}+\frac{d\bar n}{d\rho}=\bar n+1 \ .
\end{align}
The solution is given in terms of the  Bessel $I_0$ function
\begin{align}
\label{NRHO}
\bar n(\rho)=I_0(2\sqrt{\rho})-1 \ ,
\end{align}
with the correct growth in rapidity at large $y$, in  leading double-log accuracy~\cite{Kovchegov:2012mbw}.

To solve ${\cal Z}(\lambda)$ in general, we  define ${\cal Z}=e^{\cal W}$, and the equation for ${\cal W}$ now becomes
\begin{align}
\rho \frac{d^2 {\cal W}}{d\rho^2}+\frac{d{\cal W}}{d\rho}=\lambda e^{\cal W}-1 \ .
\end{align}
In terms of $u=2\sqrt{\rho}$, one has
\begin{align}
\frac{d^2{\cal W}}{du^2}+\frac{1}{u}\frac{d{\cal W}}{du}=(\lambda e^{\cal W}-1) \ ,
\end{align}
which can be written as
\begin{align}
\nabla^2 {\cal W}=(\lambda e^{\cal W}-1) \ ,
\end{align}
with $\nabla^2$ the radial part of 2-dimensional Laplacian. Note that in terms of the original generating function ${\cal Z}$, the equation is
\begin{align}
\frac{d^2{\cal Z}}{d\rho^2}=\frac{1}{{\cal Z}}\bigg(\frac{d{\cal Z}}{d\rho}\bigg)^2-\frac{1}{\rho}\frac{d{\cal Z}}{d\rho}+\frac{1}{\rho}\bigg(\lambda {\cal Z}^2-{\cal Z}\bigg) \ ,
\end{align}
which resembles  closely the third Painlevé equation (except of the last term in the last bracket).

\begin{figure}[!htb]
\includegraphics[height=7cm]{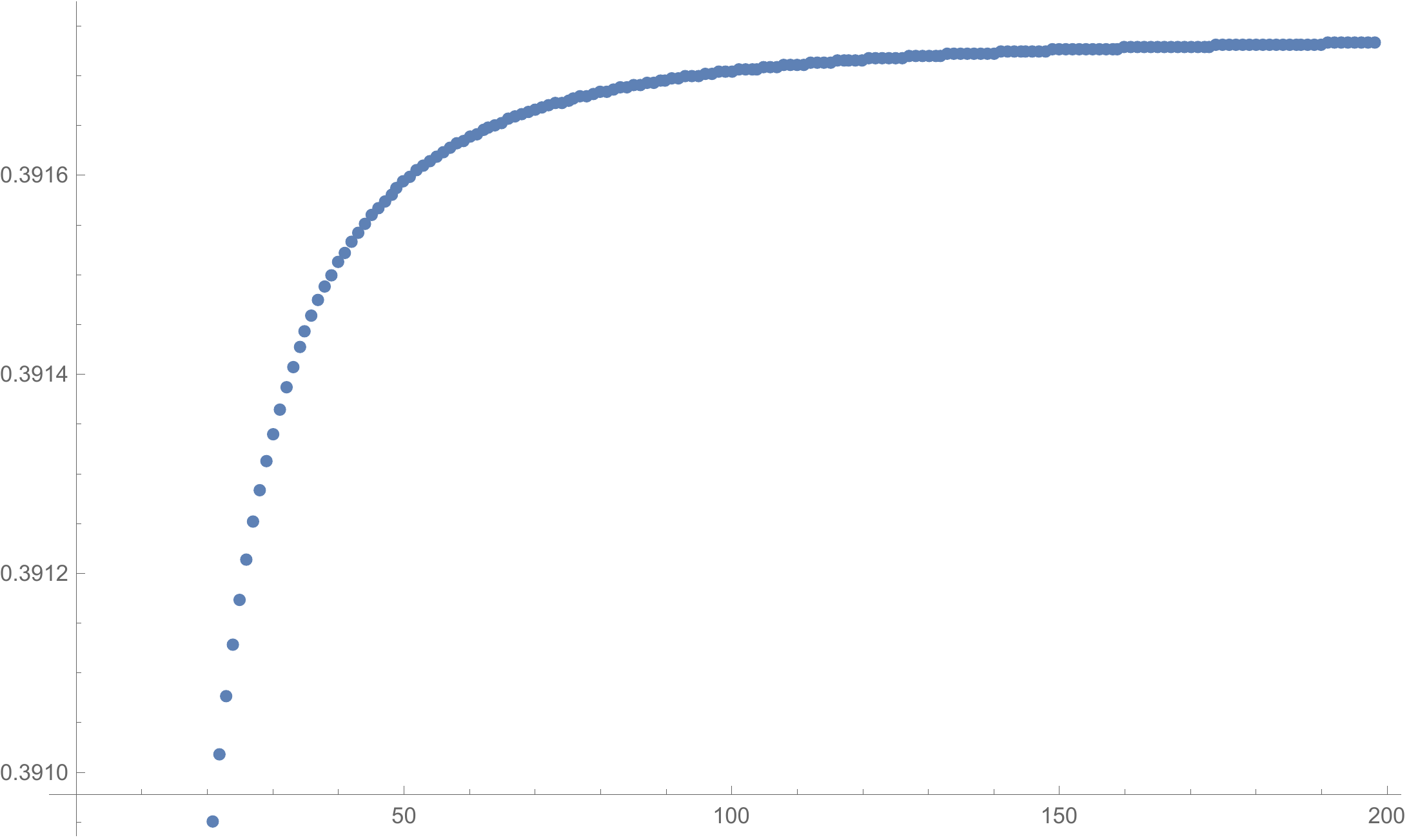}
 \caption{Behavior of the sequences  $\frac{a_{n+2}}{a_{n+1}}-\frac{a_{n+1}}{a_n}$ for $1\le n\le 198$. }
  \label{fig:diff}
\end{figure}

\subsection{Factorial moments $n_k$ in the large rapidity limit: the asymptotic moment sequence $a_n$}
To gain more insight to the multiplicity distribution, in this section we investigate property of the factorial moments
 $n_k=\frac{d^k}{d\lambda^k} {\cal Z}|_{\lambda=1}$. Successive derivatives show that they satisfy the hierarchy
\begin{align}
&\rho \frac{d^2 n_1}{d\rho^2}+\frac{d n_1}{d\rho}=n_1+1 \ , \\
&\rho\frac{d^2n_2}{d\rho^2}+\frac{dn_2}{d\rho}-n_2=2\rho(\frac{dn_1}{d\rho})^2+2n_1^2+4n_1  \ , \\
&\rho\frac{d^2n_3}{d\rho^2}+\frac{dn_3}{d\rho}-n_3=-6\rho n_1(\frac{dn_1}{d\rho})^2+6\rho\frac{dn_1}{d\rho}\frac{dn_2}{d\rho}+6n_2n_1+6n_1^2+6n_2 \ , \\
&\rho\frac{d^2n_4}{d\rho^2}+\frac{dn_4}{d\rho}-n_4=24\rho n_1^2(\frac{dn_1}{d\rho})^2-12\rho n_2(\frac{dn_1}{d\rho})^2-24\rho n_1 \frac{dn_1}{d\rho}\frac{dn_2}{d\rho}+6\rho(\frac{dn_2}{d\rho})^2 \nonumber \\
&+8\rho \frac{dn_1}{d\rho}\frac{dn_3}{d\rho}+8n_1n_3+6n_2^2+24n_2n_1+8n_3 \ ,
\end{align}
and so on. Here $n_1=\bar n$ is just the mean multiplicity. We would like to study the asymptotics of the above equations in the large $u$ limit. The solution $n_1\sim e^{2\sqrt{\rho}}\sim e^{u}$ suggests us to find the asymptotic solutions of the form $n_k\sim a_k e^{k u}$ in the large $u$ limit.

\begin{figure}[!htb]
\includegraphics[height=7cm]{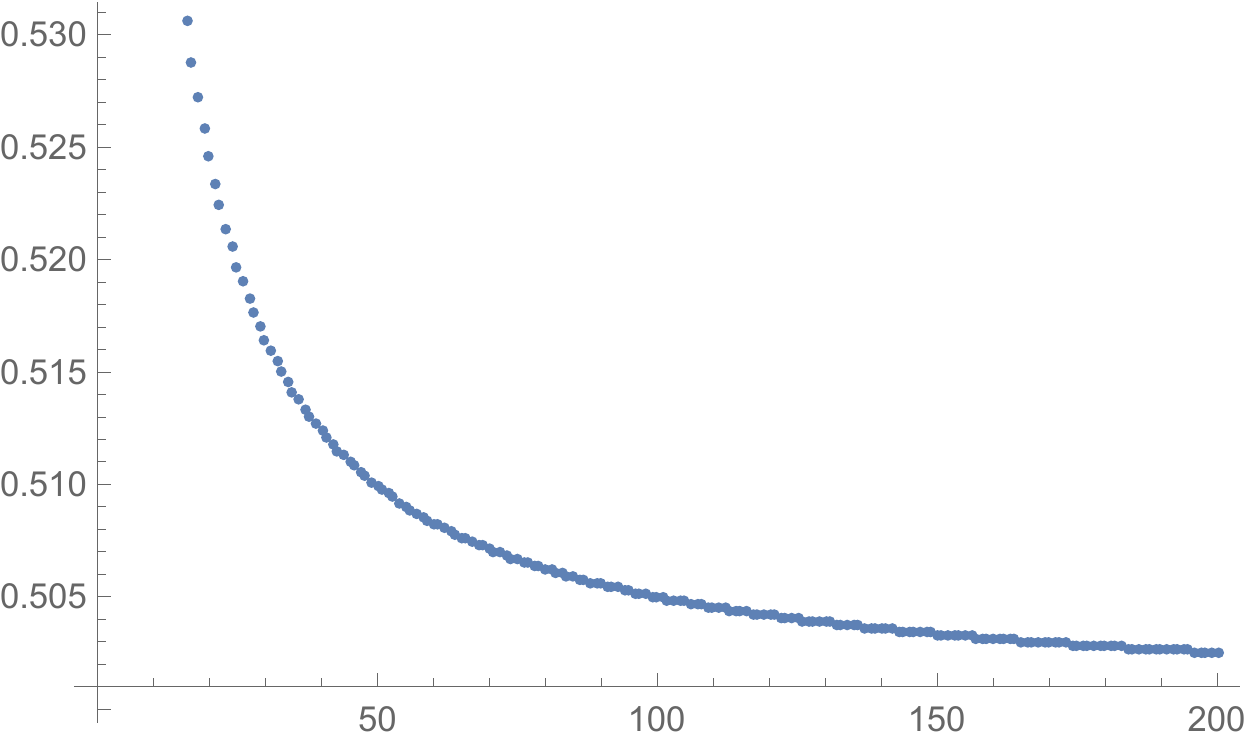}
 \caption{Behavior of the sequence ratio $\frac{(0.39174)^n(n+1)!}{a_n}$ for $1\le n\le 200$. }
  \label{fig:ratio}
\end{figure}

For this purpose, notice that in terms of $u=2\sqrt{\rho}$, the derivative terms on the left hand of equations reads
\begin{align}
\rho\frac{d^2}{d\rho^2}+\frac{d}{d\rho} \equiv \frac{d^2}{du^2}+\frac{1}{u}\frac{d}{du} \ .
\end{align}
In the large $u$ limit, we only need to  keep the $\frac{d^2}{du^2}$ term in the left-hand side and drop all terms that are exponentially suppressed on the right-hand side, to obtain the {\it leading asymptotics} of the form
\begin{align}\label{eq:defak}
n_k|_{u\rightarrow \infty} \rightarrow a_k e^{k u} \ .
\end{align}
The above defines uniquely the {\it asymptotic moment sequence} $a_n$ with the initial condition $a_0=a_1=1$.  The $a_0=1$ is due to the probability conservation while $a_1=1$ fixes the overall normalization of the sequence.
Defining
\begin{align}
b_n=\frac{a_n}{n^2} \ ,
\end{align}
it is easy to show that $b_n$ satisfies recursive relations
\begin{align}\label{eq:anrecur}
n^2b_n=B_n(b_1,b_2,...b_{n-1},b_n) \ .
\end{align}
Here the $B_n$ are the complete Bell's polynomials, which are defined for a generic sequence $b_n$ through the relation
\begin{align}\label{eq:belldef}
\exp \bigg[\sum_{n=1}^{\infty}\frac{b_nu^n}{n!}\bigg]=1+\sum_{n=1}^{\infty}\frac{B_n(b_1,...b_n)u^n}{n!} \ .
\end{align}
Notice that the Bell's polynomial satisfies the recursive relation
\begin{align}
B_{n}(b_1,....b_n)=b_n+\sum_{i=0}^{n-2}\frac{(n-1)!}{i!(n-1-i)!}B_{n-1-i}(b_1,...b_{n-1-i})b_{i+1} \ ,
\end{align}
for a generic sequence $b_n$. Given the above, Eq.~(\ref{eq:anrecur}) can be simplified to
\begin{align}
(n^2-1)b_n=\sum_{i=0}^{n-2}\frac{(n-1)!}{i!(n-1-i)!}(n-1-i)^2b_{n-1-i}b_{i+1} \ ,
\end{align}
which is can be transformed to that in Refs.~\cite{Dokshitzer:1982ia,Dokshitzer:1991wu}. Using this, the first few terms are readily generated as,
\begin{align}
&a_2=\frac{4}{3} \ , a_3=\frac{9}{4} \ , a_4=\frac{208}{45} \ , a_5=\frac{2425}{216} \ , a_6=\frac{2207}{70} \ , \nonumber \\
&a_7=\frac{1303841}{12960} \ ,a_8=\frac{3059488}{8505} \ ,a_9=\frac{7981101}{5600}\ , a_{10}= \frac{927828775}{149688}\ .
\end{align}
Furthermore, we note numerically that
\begin{align}
\label{COEFF}
\frac{a_{n+2}}{a_{n+1}}-\frac{a_{n+1}}{a_n} \rightarrow C\equiv \frac{1}{r}=0.3917 \ ,
\end{align}
when $n\rightarrow \infty$. To show this,  we display in Fig.~\ref{fig:diff} the behavior of  the sequence
$\frac{a_{n+2}}{a_{n+1}}-\frac{a_{n+1}}{a_n} $ for $n$ from $1$ to $198$. The
consecutive differences stabilize  at about $10^{-3}$ around $n=10$,  and at about $10^{-6}$ around $n=100$. The speed of convergence is around $\frac{1}{n^2}$. Asymptotically, the  asymptotic moment sequence $a_n$ are seen to approach
\begin{align}
a_n\rightarrow 2(0.39174)^n n n! \ ,
\end{align}
as illustrated in  Fig.~\ref{fig:ratio}. In the next section, we will show that the value of $C=0.3917$ in (\ref{COEFF}), is fixed by the radius of convergence of the Taylor series
for the moment sequence $a_n$.

\subsection{KNO scaling}
In this subsection we show that for large mean multiplicity $\bar n$, the distribution converges to a continuum limit as
\begin{align}
\label{PNKNO}
p_n=\frac{1}{\bar n}f(\frac{n}{\bar n}) \ ,
\end{align}
with a universal probability distribution function or scaling function $f(z)$. To show this, one introduces the re-scaled dipole number $\hat n=\frac{n}{\bar n}$. The result in the previous subsection implies that in the large rapidity limit one has for each $k$
\begin{align}
\lim_{\rho \rightarrow \infty}\bigg \langle \hat n \left(\hat n-\frac{1}{\bar n}\right)....\left(\hat n-\frac{k-1}{\bar n}\right) \bigg \rangle =a_k \ ,
\end{align}
where the expectation $\langle ...\rangle$ is defined in terms of the dipole probability distribution $p_n$ in Eq.~(\ref{eq:expect}) and $a_k$ is the asymptotic moment sequence defined in Eq~(\ref{eq:defak}).
Since $\frac{1}{\bar n}\rightarrow e^{-2\sqrt{\rho}}\rightarrow 0$ in the large $\rho$ limit, the above implies that the standard $k$-th moment of $\hat n$ converges to the asymptotic moment sequence $a_k$ as well
\begin{align}\label{eq:probconver}
\lim_{\rho \rightarrow \infty}\langle \hat n^k \rangle =a_k  \ .
\end{align}
Based on the method of moments, which states that convergence in moments implies convergence in distribution\footnote{The condition of the method of moments, namely the limiting sequence $a_k$ is a moment sequence that uniquely determines the underlining probability distribution, is indeed satisfied in this case.}, Eq.~(\ref{eq:probconver}) implies that $\hat n$ converges in the large $\rho$ limit {\it in distribution} to a random variable $z$ with the probability distribution function $f(z)$, such that the $a_k$ is nothing but its moment sequence:
\begin{align}
&\hat n=\frac{n}{\bar n}|_{\rho \rightarrow \infty}\rightarrow z, \  \ \lim_{\rho \rightarrow \infty}\langle \chi_{\hat n \in A}\rangle= P\left[z \in A\right]=\int_A f(z)dz \ , \label{eq:probiz} \\
&a_k=\int_{0}^{\infty} dz z^k f(z) \ .
\end{align}
In the above, $\chi_{\hat n \in A}$ is the characteristic function of the event $\hat n \in A$ and the expectation value $\langle.... \rangle$ is still defined with respect to $p_n$. On the other hand, $P[z\in A]$ is the probability of finding $z$ in the set $A$, according to the probability distribution $f(z)$. It is easy to see that Eq.~(\ref{eq:probiz}) is nothing but an academic way (or ``rigorous'' way) to express the KNO scaling relation Eq.~(\ref{PNKNO}).

As a result, the task reduces to find a probability distribution $f(z)$ with $a_n$ being its moment sequence. The asymptotics $a_n \rightarrow 2(0.39174)^nn (n)!$  growth quickly and satisfies the Carleman's condition, implying that the resulting $f(z)$ is unique. Unfortunately, to reconstruct $f(z)$ directly from $a_n$ is a hard inverse problem in general. However, the large $k$ asymptotics of $a_k$ already suggests
that for large $z$, the distribution decays like
\begin{align}
f(z)|_{z\rightarrow \infty}\rightarrow 2r^2 ze^{-zr} \ .
\end{align}
This will be confirmed by different arguments in the following section.

\begin{figure}[!htb]
\includegraphics[height=8cm]{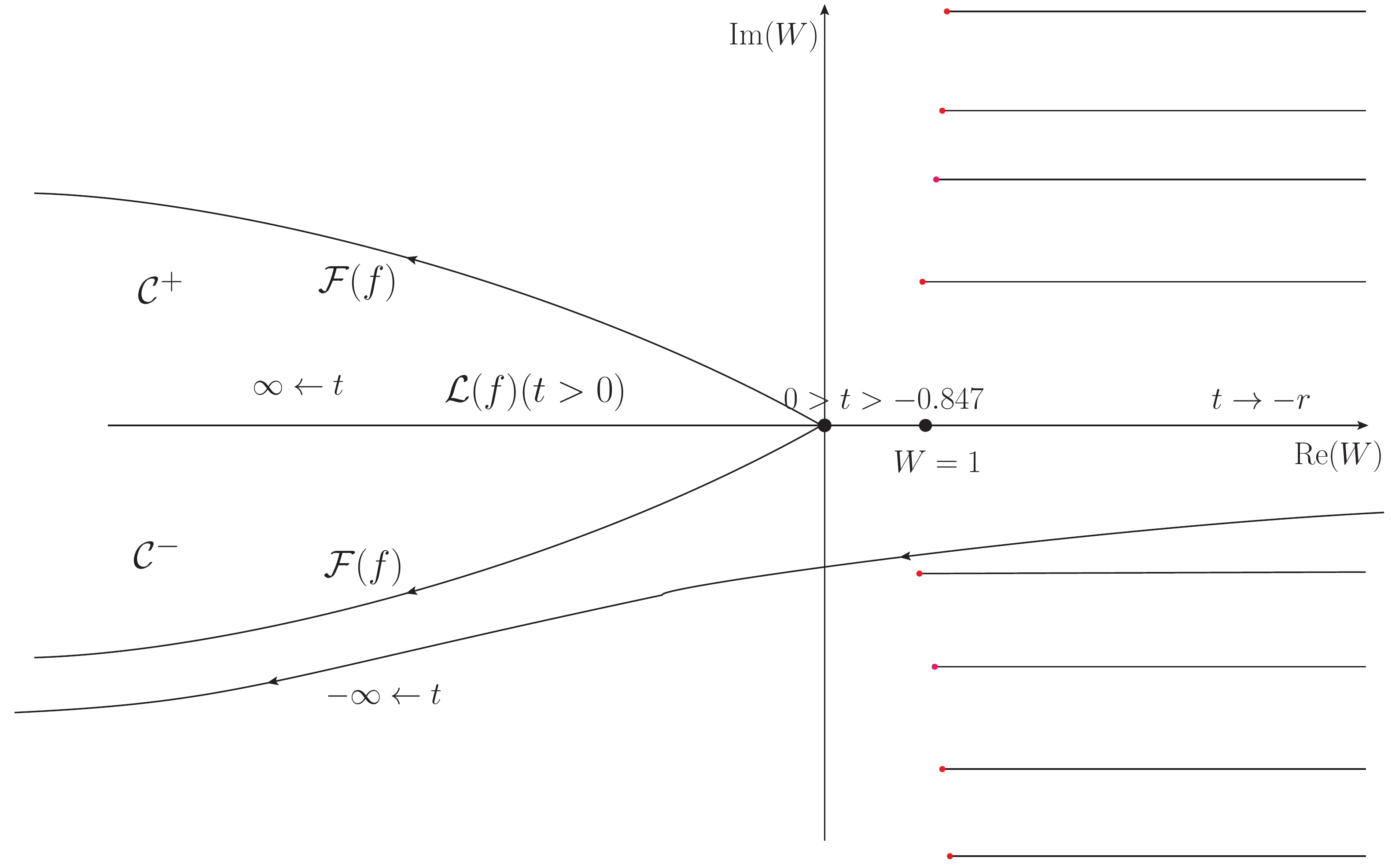}
 \caption{We display different regions for the analytic continuation for $W$. Explicitly shown are the trajectories for end points of $W(t)$ in~(\ref{eq:Winte}), whereas the integration path is from $U=1$ to $W$ without touching singularities.  The branch cuts are shown as horizontal lines. The labels  ${\cal F}(f)$, ${\cal L}(f)$ denote the Fourier and Laplace transforms of the underlining probability distribution $f$. In particular, for $t>0$, the end-point $W$ must locate at the negative real axis, while the path is from $U=1$ to $U=-|W|$ along the real axis, with the pole at $U=0$ circumvented through a small circle in the half upper or lower plane, which gives rise to the imaginary part $\pm i\pi$ for $u$. On the other hand, to get the Fourier transform, $u$ must have an imaginary part $\pm \frac{i\pi}{2}$. For this the endpoint $W$ must locate along the curve ${\cal F}(f)$, while the integration path is first from $U=+1$ to $U=-{\rm |Re W|}$ along real axis as before, and than vertically to ${\rm Im }W$. In particular, the $\frac{\pi i}{2}$ goes into the upper half plane,  and the $-\frac{\pi i}{2}$ goes into the lower half plane. The region ${\cal C}^{+}$ and ${\cal C}^{-}$ between the Laplace transform and the Fourier transform, is  the usual analyticity region for the  characteristic function of the a positively supported probability distribution. }
  \label{fig:analy}
\end{figure}

\section{The KNO scaling function}
\label{SEC_KNO}
Generally, reconstructing a probability distribution from its moment sequence is a hard inverse problem. However, in the current case, since the asymptotic moment sequence $a_n$ is inherited from a second order differential equation in the large $\rho$ limit, we expect that its exponential generating function $Z(t)$
\begin{align}
Z(t=-e^{u})=\sum_{n=0}^{\infty}\frac{a_n}{n!}e^{nu} \ ,
\end{align}
also satisfies a second order differential equation. Indeed, if one introduces
\begin{align}
W=\ln Z \ ,
\end{align}
then it is not hard to show using Eqs.~(\ref{eq:anrecur}),~(\ref{eq:belldef}) that $W$ obeys a simpler equation\footnote{An equivalent equation has been obtained before for jet multiplicities~\cite{Dokshitzer:1982ia,Dokshitzer:1991wu}.}
\begin{align}\label{eq:differW}
\frac{d^2}{du^2}W(u)=e^{W(u)}-1 \ ,
\end{align}
with the boundary condition that for $u\rightarrow -\infty$, $W \rightarrow e^{u}$. (\ref{eq:differW}) can then be integrated readily,
 and analytically continued to the whole complex plane to obtain the Fourier-Laplace transform of the scaling function $f$
\begin{align}
Z(t)={\cal L}(f)(t)=\int_{0}^{\infty} e^{-tz}f(z)dz \ .
\end{align}
The complex-analytic Fourier-Laplace transform automatically includes the standard Fourier-transform of $f$ at $t= i {\cal R}$, or $u= {\cal R}\pm \frac{\pi}{2}i $, from which
$f(z)$ can be finally obtained by taking the inverse Fourier transform. We should emphasize that the complex analytic methods are crucial to obtain the Fourier transform of $f(z)$,  since its Taylor expansion in terms of the moments
\begin{align}\label{eq:serierW}
Z(it)=\sum_{n=0}^{\infty}\frac{a_n}{n!}(-1)^n(it)^n \ ,
\end{align}
has a finite radius of convergence $r=\frac{1}{C}$.

\subsection{Integral representation of $W$}
To proceed, we first notice that  (\ref{eq:differW}) can be integrated as~\cite{Dokshitzer:1982ia,Dokshitzer:1991wu}
\begin{align}
u=c+\int_{1}^{W}\frac{dU}{\sqrt{2e^{U}-2U-2}} \ ,
\end{align}
where the constant $c$ can be fixed by imposing the boundary condition $W|_{u\rightarrow -\infty}\rightarrow e^{u}$ as
\begin{align}
c=-\int_{1}^{0}dU \bigg(\frac{1}{\sqrt{2e^U-2U-2}}-\frac{1}{U}\bigg) \ .
\end{align}
Therefore we obtain the integral representation of $W$
\begin{align} \label{eq:Winte}
u=\int_{0}^{1}dU \bigg(\frac{1}{\sqrt{2e^U-2U-2}}-\frac{1}{U}\bigg)+\int_{1}^{W(u)}\frac{dU}{\sqrt{2e^{U}-2U-2}} \ .
\end{align}
The integral is along the real axis from $1$ to $W(u)$. This completely determines $W(u)$ for $-\infty<u<\ln r$, where as $u \rightarrow\ln r^-$ from the lower side,  $W$ diverges to positive infinity. This corresponds to  nothing but the radius of convergence $r=1/C$ for the series representation (\ref{eq:serierW}),
\begin{align}\label{eq:logr}
\ln r=\int_{0}^{1}dU \bigg(\frac{1}{\sqrt{2e^U-2U-2}}-\frac{1}{U}\bigg)+\int_{1}^{\infty}\frac{dU}{\sqrt{2e^{U}-2U-2}}= 0.93715 \ ,
\end{align}
with
\begin{align}
C=\frac{1}{r}=e^{-0.93713}=0.391743 \ .
\end{align}
This value of $C$ is in good agreement with the numerical observation (\ref{COEFF}), since the convergence radius is nothing but inverse to the rate of the exponential part for $a_n$ when $n\rightarrow \infty$.

\begin{figure}[!htb]
\includegraphics[height=8cm]{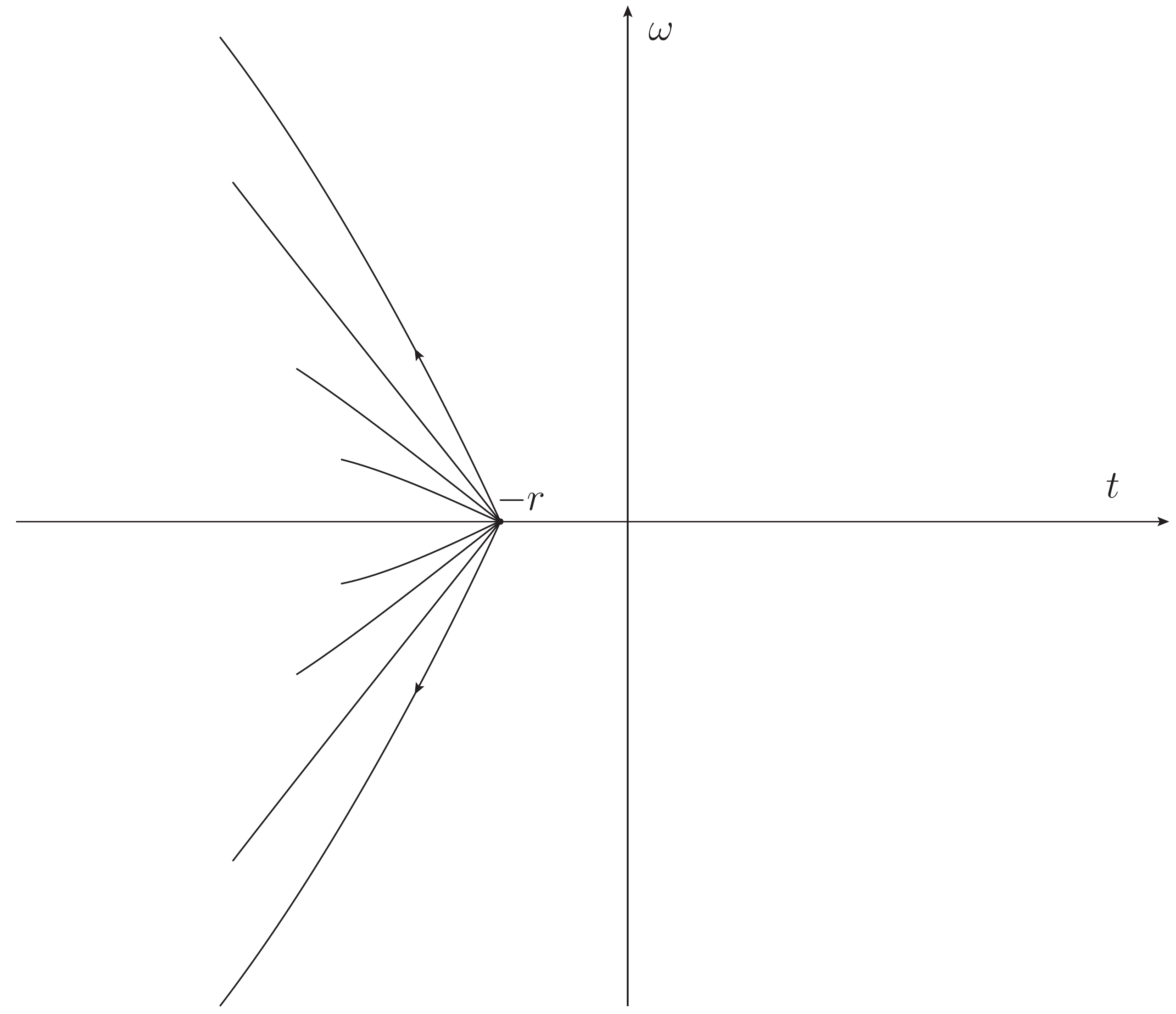}
 \caption{The analyticity structure of the Fourier-Laplace transform of $f$. The Laplace transform corresponds to the positive real axis, while the Fourier transform corresponds to the imaginary axis. The singularity at $t=-r$ controls the large $z$ behavior of the probability distribution, and is the source of infinitely many branch cuts corresponding to the roots. The outer ones shown with arrow correspond to $w_1$ and $\bar w_1$.}
  \label{fig:analyf}
\end{figure}

Furthermore, the behavior of $W$ when $u\rightarrow \ln r^-$ can be worked out  from the equation
\begin{align}
\int_{W}^{\infty}\frac{dU}{\sqrt{2e^{U}-2U-2}}=\ln r-u \ ,
\end{align}
which after expanding the square-root around the dominant term $2e^{U}$, reads
\bea
\sqrt{2}e^{-\frac{W}{2}}+\frac{1}{3\sqrt{2}}e^{-\frac{3}{2}W}(W+\frac{5}{3})+{\cal O}(W^2e^{-\frac{5}{2}W})=\ln r-u \ .
\eea
To solve it,  we split $W=W_0+W_1$  with
\bea
e^{-\frac{W_0}{2}}&=&\frac{\ln r-u}{\sqrt{2}} \ , \nonumber\\
e^{-\frac{W_1}{2}}&=&1-\frac{(\ln r-u)^2}{12}\bigg(-2\ln \frac{\ln r-u}{\sqrt{2}}+\frac{5}{3}\bigg) \ ,
\eea
which implies for $Z$ the following
\begin{align}\label{eq:asymWpole}
Z(u)=\frac{2}{(\ln r-u)^2}+\frac{1}{3}\bigg(-2\ln \frac{\ln r-u}{\sqrt{2}}+\frac{5}{3}\bigg)+{\cal O}(\ln r-u)  \ .
\end{align}
Therefore $Z$ has a double ``pole'' around the singularity, which exactly implies that $f(z)\sim 2r^2ze^{-rz}$ at large $z$. Moreover, the coefficient $2$ matches precisely the $2$ in the asymptotic form of $a_k$. This has to be compared to the $0+1$ reduction case, where one has the equation
\begin{align}
u=\int_{0}^{1}dU \bigg(\frac{1}{e^{U}-1}-\frac{1}{U}\bigg)+\int_{1}^{W(u)}\frac{dU}{e^U-1} \ ,
\end{align}
and as $u\rightarrow 0$,
\begin{align}
Z(u)\sim -\frac{1}{u} \ ,
\end{align}
which implies that $Z$ has only a single pole.

\subsection{Analyticity structure of $Z$}
We have already obtained the $W(t)$ in the region $-r<t<0$. It is time to extend it to the whole complex plane. For this, we need to understand the singularity structure
of the integrand
\begin{align}
G(w)=\frac{1}{\sqrt{2e^w-2w-2}} \ .
\end{align}
The entire function $g(w)=2e^w-2w-2$ has a double pole at $w=0$, and infinitely many non-zero single poles $w_n$ and $\bar w_n$. It admits  the infinite product expansion
\begin{align}
2e^w-2w-2=w^2e^{\frac{w}{3}}\prod_{n=1}^{\infty}\frac{(w-w_n)(w-\bar w_n)}{|w_n|^2}e^{\frac{w}{w_n}+\frac{w}{\bar w_n}} \ ,
\end{align}
One can show that all the non-zero poles are in the right half-plane, and the first root occurs at $w_1=2.088+7.46149 i$, $\bar w_1 =2.088-7.46149i$. For large $n$, we have  $w_n\rightarrow \ln (2n+\frac{1}{2})\pi+ (2n+\frac{1}{2})\pi i$. In fact, by expanding around $w=(2n+\frac{1}{2})\pi i\equiv A_n i$, one obtains an approximate formula for the location of the poles
\bea
w_n&=&x_n+iy_n \ , \nonumber\\
x_n&=&\ln A_n+\frac{\ln^2A_n-1}{2A_n^2}+{\cal O}\bigg(\frac{\ln^4 A_n}{A_n^4}\bigg) \ ,\nonumber \\
y_n&=&A_n-\frac{1+\ln A_n}{A_n}+{\cal O}\bigg(\frac{\ln^3 A_n}{A_n^3}\bigg) \ .
\eea
This approximation is better than expected, as the first pole is already reproduced within three digits accuracy. Given the poles, we define $G(w)$ as
\begin{align}
G(w)=\frac{1}{w}e^{-\frac{w}{6}}\prod_{n=1}^{\infty}\frac{-|w_n|}{(w-w_n)^{\frac{1}{2}}(w-\bar w_n)^{1/2}}e^{-\frac{w}{2w_n}-\frac{w}{2\bar w_n}} \ .
\end{align}
The square roots are defined with branch cuts extending from $w_n$ to $w_n+\infty$, and $\bar w_n$ to $\bar w_n+\infty$, namely, for $w_n$ the arguments go from $0$ to $2\pi$ and the same for $\bar w_n$. The minus sign in the numerator will guarantee that $G(w)$ is positive along the positive real axis,  while negative along negative real axis. Therefore, $G(w)$ is analytic in the left half-plane ${\rm Re}(w)<0$ and has a single pole at $w=0$. In the right half-plane it has infinitely many branch cuts extending to positive infinity.

Given the knowledge of $G(w)$ and its singularity structure, one can determine
$Z(t)$ for $t$ outside the initial region $-r<t<0$ by specifying the integration paths (in fact, the endpoint) for $W$. We first demonstrate this for $-\infty<t<-r$. One first notices that for $u\rightarrow \ln r^+$ or $t\rightarrow -r^-$, $W$ must go from $\infty\pm 2\pi i$. In fact, the real and imaginary part for $W$ when $u$ is still real,  must satisfy
\begin{align}\label{eq:analyW1}
\int_{{\rm Re}(W)-i{\rm Im}(W)}^{{\rm Re}(W)+i{\rm Im}(W)} dw G(w)=0 \ .
\end{align}
Therefore, we need to show that asymptotically ${\rm Im}(W) \rightarrow 2\pi$, as ${\rm Re}(W)\rightarrow +\infty$. Indeed, this is the case, since for very large ${\rm Re}(W)$, we have
\begin{align}
\int_{{\rm Re}(W)-i{\rm Im}(W)}^{{\rm Re}(W)+i{\rm Im}(W)}dw G(w)\sim \int_{{\rm Re}(W)-i{\rm Im}(W)}^{{\rm Re}(W)+i{\rm Im}(W)}dw e^{-w/2} \ ,
\end{align}
for which ${\rm Im}(W) \rightarrow 2\pi$ is justified. When the condition in (\ref{eq:analyW1}) is satisfied, $u$ as  given by (\ref{eq:Winte}), is  real and larger then $\ln r$. Furthermore, when $u$ increases, the real part of $W$ decreases, and one can show that ${\rm Im} W$ has to increase. However, the path will never meet the branch cuts for $G(z)$, and as $t\rightarrow -\infty$ or $u\rightarrow +\infty$, ${\rm Re} (W)$ approaches to $-\infty$ along the curve depicted in Fig.~\ref{fig:analy}.

Similarly, the integration paths for the Fourier transform ${\cal F}(f)$, corresponding to ${\rm Im}(u)=\pm \frac{\pi}{2}$ and the Laplace transform with $t>0$, corresponding to ${\rm Im }(u)=\pm \pi$, can be worked out. The results are shown in Fig.~\ref{fig:analy}. In particular, the pole of $G(w)$ at $w=0$ contributes to the desired imaginary part $\pm i\pi$ in case of the Laplace transform when the endpoint $W$ for the integration path moves from the positive to negative axis. Similarly, the aggregation of the imaginary parts $\pm i\pi$ due to the pole and another $\mp i\frac{\pi}{2}$ acquired in the vertical path when $U$ goes from $({\rm Re}(W),0)$ to $({\rm Re}(W),{\rm Im}(W))$ contributes to the total $\pm i\frac{\pi}{2}$ in case for the Fourier transform.  The analyticity structure for the Fourier-Laplace transform is summarized in Fig.~\ref{fig:analyf}. There is clearly a one to one correspondence between Fig.~\ref{fig:analy} and Fig.~\ref{fig:analyf}.

\subsection{Asymptotics of the KNO scaling function}
Given the analytic Fourier-Laplace transform, the behavior of the scaling function $f(z)$ follows readily. In fact, the asymptotics of $f(z)$ for large $z$ is closely related to the behavior of $Z(t)$ around $t=-r$, while the small-$z$ behavior of $f(z)$ can be deduced from the large ${\rm Re} (t) \rightarrow +\infty$ asymptotics for $Z(t)$. We discuss them separately.
\\
\\
{\bf Small $z$ behavior:}
\\
To determine the decay rate at small $z$, one needs to work out the behavior for ${\cal L}(f)(t)$ at large $t$ or ${\cal F}(f)(\omega)$ at large $\omega$. It is sufficient to consider the Laplace transform. Clearly, for $t\rightarrow \infty$, ${\rm Re}(W)$ must goes to $-\infty$. More precisely, the Laplace transform is determined by the representation
\begin{align}
\ln t=\int_{-1}^{0}dU \bigg(\frac{1}{\sqrt{2e^U-2U-2}}+\frac{1}{U}\bigg)+\int_{W(t)}^{-1}\frac{dU}{\sqrt{2e^{U}-2U-2}}  \ ,
\end{align}
which will guarantee that for $t \rightarrow 0^+$, one has the correct boundary condition.
\begin{align}
W(t) \rightarrow -t \ .
\end{align}
For large $|W|$ we can expand,
\begin{align}
\int_{1}^{|W|}\frac{dU}{\sqrt{2U-2+2e^{-U}}}=\int_{1}^{|W|}\frac{dU}{\sqrt{2U-2}}+\int_{1}^{|W|}\bigg(\frac{1}{\sqrt{2U-2+2e^{-U}}}-\frac{1}{\sqrt{2U-2}}\bigg) \ .
\end{align}
Now for large $|W|$ the last integral is convergent, so that  for $t \rightarrow \infty$
\begin{align}
\ln t=\int_{-1}^{0}dU \bigg(\frac{1}{\sqrt{2e^U-2U-2}}+\frac{1}{U}\bigg)+\int_{1}^{\infty}\bigg(\frac{1}{\sqrt{2U-2+2e^{-U}}}-\frac{1}{\sqrt{2U-2}}\bigg)+\sqrt{-2W-2} \ ,
\end{align}
or
\begin{align}
\ln t +0.411926=\sqrt{-2W-2}  \ , \qquad\rightarrow\qquad
W=-1-\frac{1}{2}\ln^2 (\alpha t)+{\cal O}(e^{-\frac{1}{2}\ln^2 (\alpha t)}) \ ,
\end{align}
which implies that for large $t$ \ ,
\begin{align}
{\cal L}(f)(t)\rightarrow \exp \bigg[-1-\frac{1}{2}\ln^2 (\alpha t)\bigg] \ ,
\end{align}
where $\alpha=1.50972$ is a pure number. It is not hard to show that above gives the correct asymptotics for $Z(t)$ with ${\rm Re}(t) \rightarrow +\infty$.

Given the above, at small $z$ one can simply shift the contour of inverse Fourier transform to $\frac{1}{z}+it$ in order to reach the asymptotic region. After simple algebra, one has
\begin{align}
f(z)=\frac{1}{2\pi z}\exp\bigg(-\frac{1}{2}\ln^2 \frac{\alpha}{z}\bigg)\int_{-\infty}^{\infty} dt \exp \bigg[-\ln \frac{\alpha}{z}\ln(1+it)+it-\frac{1}{2}\ln^2(1+it)\bigg] \ .
\end{align}
One must now determine the asymptotics of the integral at small $z$. Applying the saddle point analysis one finally has
\begin{align}
f(z)=\frac{1}{z}\ln \frac{\alpha}{z}\exp\bigg(-\frac{1}{2}\ln^2 \frac{\alpha}{z}-\ln \frac{\alpha}{z} \ln \ln \frac{\alpha}{z}+\ln \frac{\alpha}{z}-\frac{1}{2}\ln^2 \ln \frac{\alpha}{z}-2+{\cal O}(\frac{\ln^2 \ln \frac{\alpha}{z}}{\ln \frac{\alpha}{z}})\bigg) \ .
\label{ysmally}
\end{align}
The speed of growth is much slower than the $0+1$D reduction, but much faster than the $1+2$D case.
\\
\\
{\bf Large $z$ behavior:}
\\
To determine the large $z$ behavior, we now use the Fourier inverse-transform
\begin{align}
f(z)=-i\int_{-i\infty}^{i\infty} \frac{dt}{2\pi}e^{tz}Z(t) \ .
\end{align}
For large $z$, one would like to shift the integration path as left as possible. The singularity at $t=-r$ will prevent shifting the contour furthermore, and gives rise to the leading exponential decay. The singular part of $Z$ near $t=-r$ along the real axis, has already been given in Eq.~(\ref{eq:asymWpole}) in term of $u=\ln (-t)$. Expressed in  terms of $t$, it reads
\begin{align}
Z(t)|_{t\rightarrow -r}=\frac{2r^2}{(t+r)^2}-\frac{2r}{t+r}+\frac{1}{3}\bigg(-2\ln \frac{t+r}{\sqrt{2}r}+\frac{13}{6}\bigg)+{\cal O}((t+r)\ln (t+r)) \ .
\end{align}
The above applies to all the directions $t+r=|t+r|e^{i\theta}$ when $|\theta|\le \frac{\pi}{2}$, including the vertical line $t=-r+ix$
\begin{align}
Z(-r+ix)|_{x\rightarrow 0}=-\frac{2r^2}{x^2}-\frac{2r}{ix}+\frac{1}{3}\bigg(-2\ln \frac{ix}{\sqrt{2}r}+\frac{13}{6}\bigg)+{\cal O}(x\ln (x)) \ .
\end{align}
Now, shifting the contour with a small circle centered at $t=-r$, we have
\begin{align}
f(z)e^{rz}=\int_{-\infty}^{-\epsilon}\frac{dx}{2\pi}e^{ix z}Z(-r+ix)+\int_{\epsilon}^{\infty}\frac{dx}{2\pi}e^{ix z}Z(-r+ix)+\int_{C_{\epsilon}}\frac{dx}{2\pi}e^{ix z}Z(-r+ix) \ .
\end{align}
Using the explicit form of the singularity for $Z(-r+ix)$ for small $x$, and the fact that $Z(-r+ix)$ decays as $e^{-\frac{\ln^2 ix}{2}}$ for large $ix$,
 and is infinitely smooth when $x\ne 0$,  we obtain
\bea
f(z)=2r\bigg(rz-1+{\cal O}(\frac{\ln z}{z})\bigg)e^{-rz} \ , \ z \rightarrow \infty \  . \label{eq:fsamylarge}
\eea
Eqs. (\ref{ysmally}) and (\ref{eq:fsamylarge}) are the major results of this section.

\begin{figure}[!htb]
\includegraphics[height=9cm]{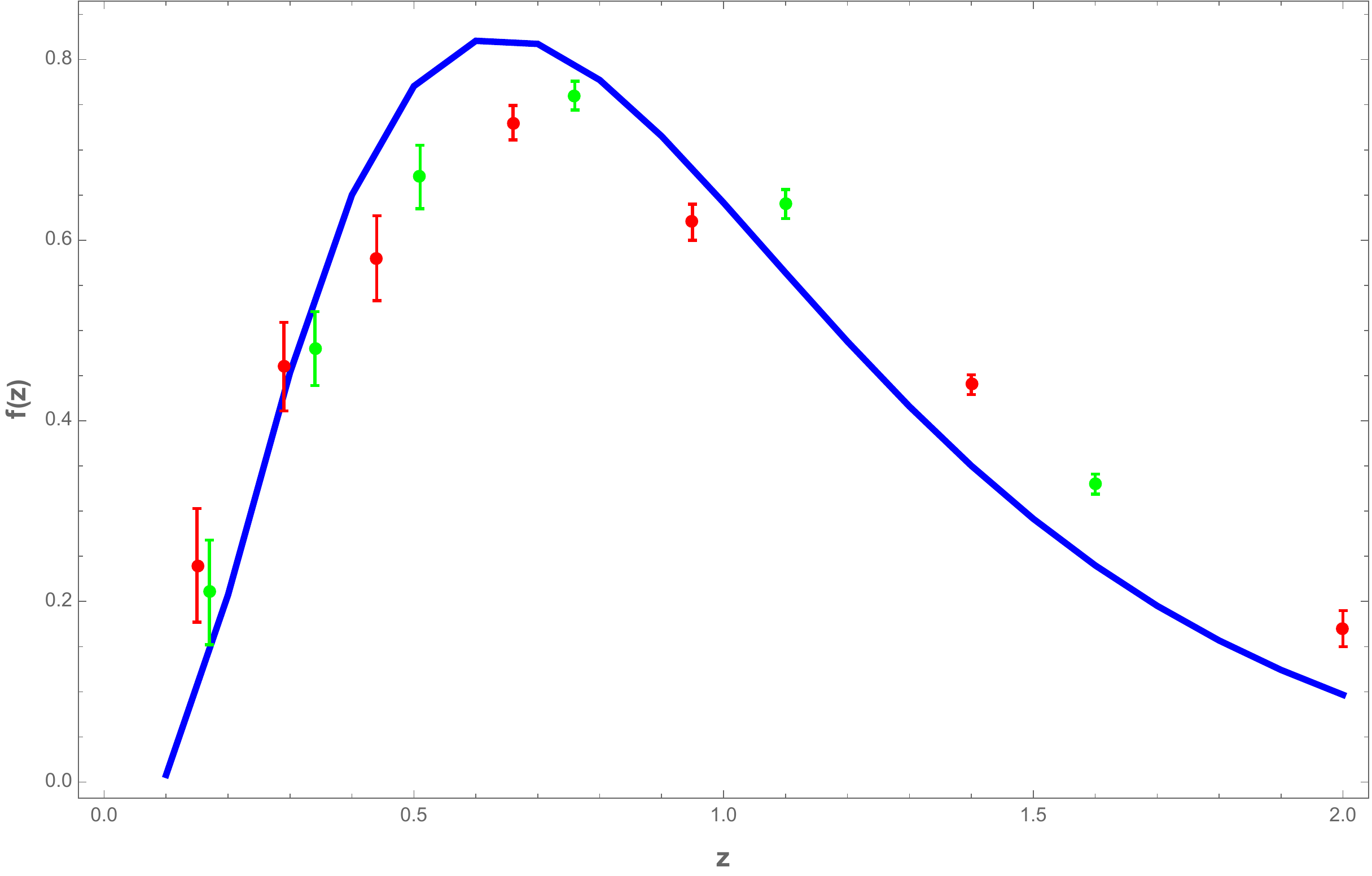}
 \caption{The exact scaling of the KNO particle multiplicity $f(z)$ as in (\ref{FZALL}) with a peak around $z=0.6$, following from the DLA  of Mueller$^\prime$s dipole wave-function evolution (solid-blue). The data is the measured KNO scaling function $\Psi(z)$  for the charged particle multiplicities in ep DIS scattering, as reported by the H1 collaboration in~\cite{H1:2020zpd},
 for $\sqrt{s}=319 \,{\rm GeV}$,
 photon virtualities $40<Q^2<100\, {\rm GeV}^2$ and charged particle pseudorapidities in the range $0<\eta<4$, for two different
 inelasticities $0.15<y_I<0.3$ (green) and $0.3<y_I<0.6$ (red)~\cite{H1:2020zpd}.}
  \label{fig:scal}
\end{figure}

\begin{figure}[!htb]
\includegraphics[height=9cm]{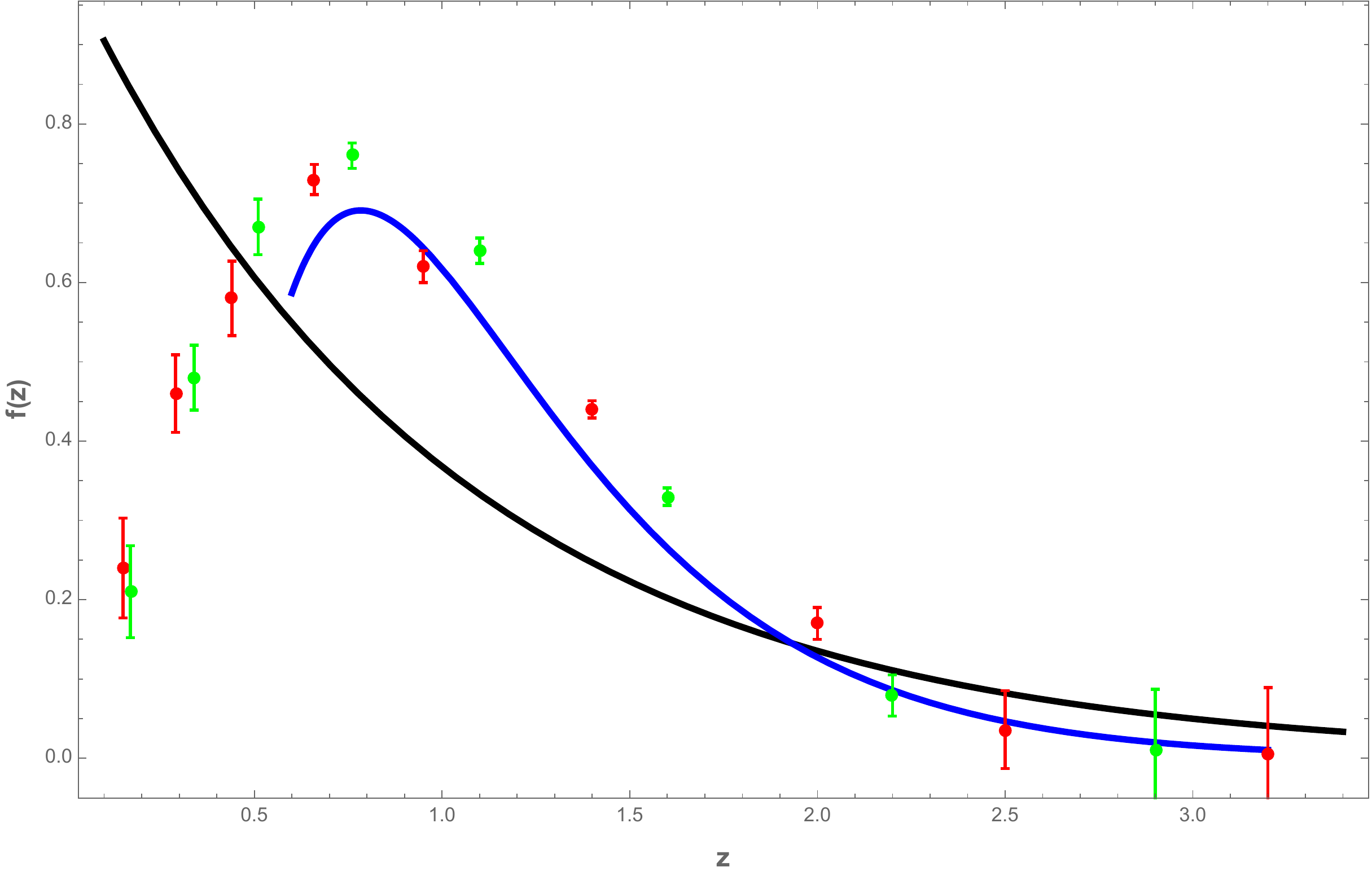}
 \caption{The exact asymptotic scaling of the KNO particle multiplicity $f(z)$ as in (\ref{eq:fsamylarge}), following from the DLA of Mueller$^\prime$s dipole wave-function evolution (solid-blue). For comparison we show the KNO particle multiplicity $e^{-z}$, following from the $0+1$D reduction or diffusive approximation of  Mueller$^\prime$s dipole wave-function evolution (solid-black).
The data are from the H1 collaboration in~\cite{H1:2020zpd}.}
  \label{fig:asymp}
\end{figure}

\section{DLA KNO scaling versus  H1 data}
\label{SEC_H1}
Although the asymptotics of the DLA for $f(z)$ can be obtained exactly as presented in the previous section, the full shape of $f(z)$ can only be obtained numerically,
with the help of the inverse Fourier transform. For that, we choose to invert with the complex valued path
\begin{align}
{\cal C}_{\gamma}=-x\pm \gamma\pi \sqrt{2x}i \ , 0<x<\infty \ ,
\end{align}
in terms of which the  inverse Fourier transform reads
\bea
\label{FZALL}
f(z)&=&-\frac{i}{2\pi}\int_{{\cal C}_{\gamma}}\frac{WdW}{\sqrt{2e^W-2W-2}}\exp \bigg[g(W)+W-zWe^{g(W)}\bigg] \ , \nonumber\\
g(W)&=&\int_{W}^{0}dU\bigg(\frac{1}{\sqrt{2e^{U}-2U-2}}+\frac{1}{U}\bigg) \ .
\eea
This choice of the path guarantees that the integrand decays exponentially for large $x$, for the parameter $\gamma$ in the range $\frac{1}{\sqrt{2}}\le \gamma\le 2$.
In fact, the natural path for the Fourier transform asymptotically approaches  ${\cal C}_{\frac{1}{\sqrt{2}}}$.  Clearly, the result is path independent,
provided that the path insures convergence at infinity. For convenience, the numerical values of $f(z)$ in the non-asymptotic regime are tabulated in
Table~\ref{tab:scaling}.
\begin{table*}
\caption{\label{tab:scaling} Table of the scaling function }
\begin{tabular}{ c|c|c|c}
\hline
$z$ & $f(z)$&$z$&$f(z)$ \\
\hline\hline
0.1&0.01&1.1& 0.56  \\
\hline
0.2& 0.21&1.2& 0.49  \\
\hline
0.3& 0.45&1.3& 0.42   \\
\hline
0.4& 0.65&1.4& 0.35   \\
\hline
0.5& 0.77&1.5& 0.29   \\
\hline
0.6&0.82&1.6&  0.24   \\
\hline
0.7& 0.82&1.7& 0.20    \\
\hline
0.8& 0.78&1.8& 0.16    \\
\hline
0.9& 0.72&1.9& 0.12    \\
\hline
1.0& 0.64&2.0& 0.1    \\
\hline
\end{tabular}
\end{table*}

In Fig.~\ref{fig:scal} we show in solid-blue, the numerical result for $f(z)$ (\ref{FZALL}) in the range $0.1<z<2$, with a peak around $z=0.6$.
The result for the DLA $f(z)$ compares well to the  measured KNO scaling function $\Psi(z)$ for the charged particle multiplicities,
reported by the H1 collaboration~\cite{H1:2020zpd}. The charged multiplicities are measured in ep DIS scattering, for $\sqrt{s}=319 \,{\rm GeV}$,
 photon virtualities $40<Q^2<100\, {\rm GeV}^2$ and charged particle pseudorapidities in the range $0<\eta<4$, for two different
 inelasticities $0.15<y_I<0.3$ (green) and $0.3<y_I<0.6$ (red). In Fig.~\ref{fig:asymp} we compare the KNO scaling
 function in the diffusive regime $e^{-z}$  (solid-black), to the exact DLA asymptotics  (\ref{eq:fsamylarge})
 (solid-blue). The DIS data at HERA support the DLA solution we have presented
 for all range of $z$, over the diffusive solution.

\section{Relation to jet evolution}\label{SEC_BMS}
As we mentioned in our introductory remarks, the DLA scaling function for the Mueller's dipole evolution,  is identical to that of  jet evolution~\cite{Dokshitzer:1982ia,Dokshitzer:1991wu}\footnote{The equality has been briefly pointed out in~\cite{Liou:2016mfr}.}. This is not accidental, and follows from the correspondence between the BK-BMS evolution equation~\cite{Banfi:2002hw,Weigert:2003mm,Marchesini:2003nh,Hatta:2013iba,Caron-Huot:2015bja}, as we now detail. When limited to the virtual part only, this correspondence is also called ``soft-rapidity'' correspondence~\cite{Vladimirov:2016dll}. We will provide a pedagogical introduction to this correspondence at leading order, using the generating functional approach in~\cite{Mueller:1993rr}, by emphasizing the underlining time and energy ordering of the soft gluon emission. In particular, we will show that the double logarithm limit in the BMS equation has the same angular ordering as that in~\cite{Dokshitzer:1982ia,Dokshitzer:1991wu}, which maps to the dipole size ordering of the dipole evolution under the conformal transformation discussed in~\cite{Cornalba:2007fs,Vladimirov:2016dll}. As a result, the two scaling functions are identical.

\subsection{Generating function approach to BMS equation}
\begin{figure}[!htb]
\includegraphics[height=5cm]{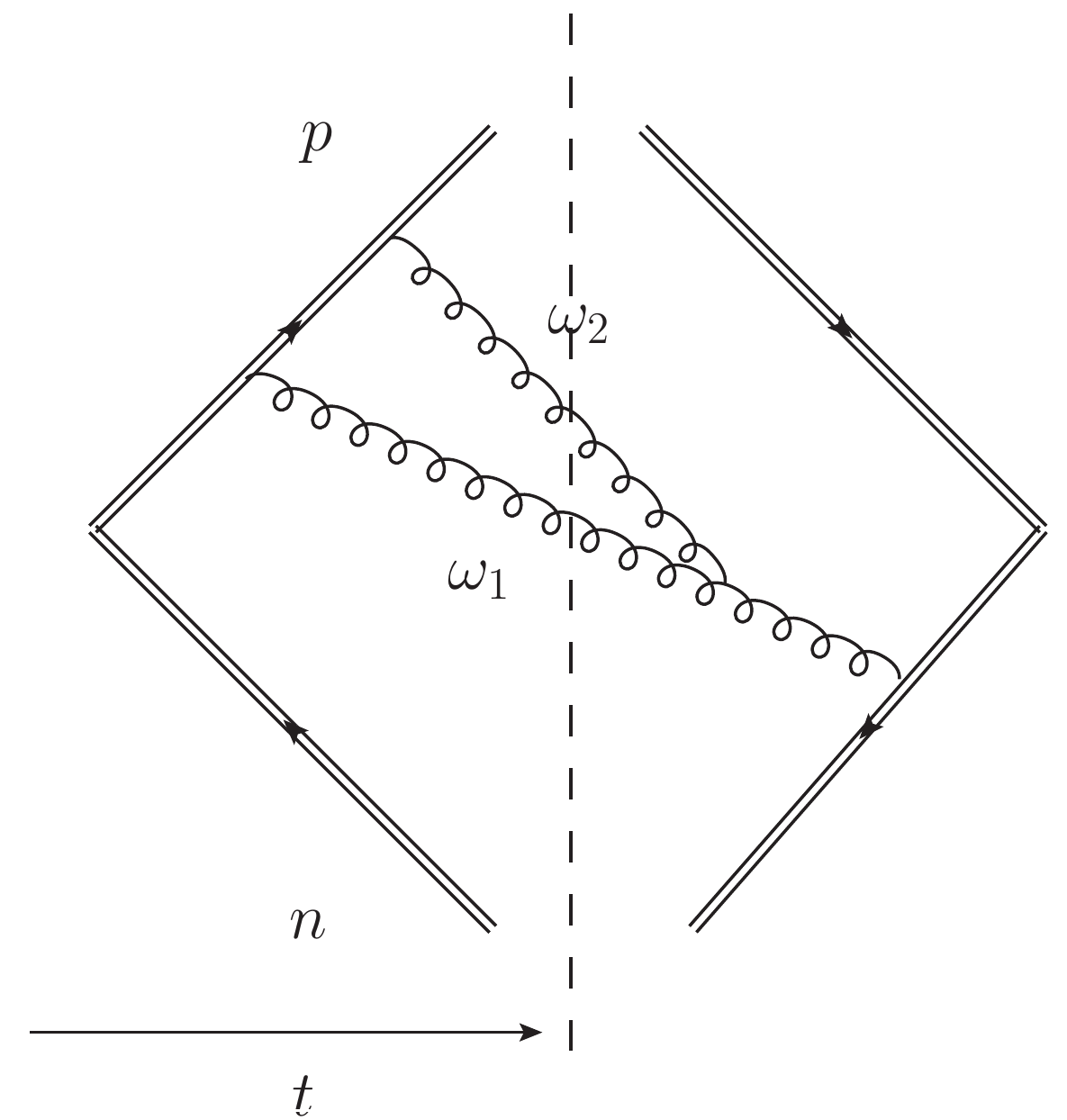}
 \caption{A sample diagram with two real gluons. The direction of the time is shown as the arrow for the amplitudes (left to the cut), and opposite for the conjugate amplitudes (right to the cut). The region $\omega_1\gg \omega_2$ contributes to leading logarithm $\alpha_s^2\ln^2 \frac{E}{E_0}$ for $\sigma_2$.}
  \label{fig:2gluonbms1}
\end{figure}

\begin{figure}[!htb]
\includegraphics[height=5cm]{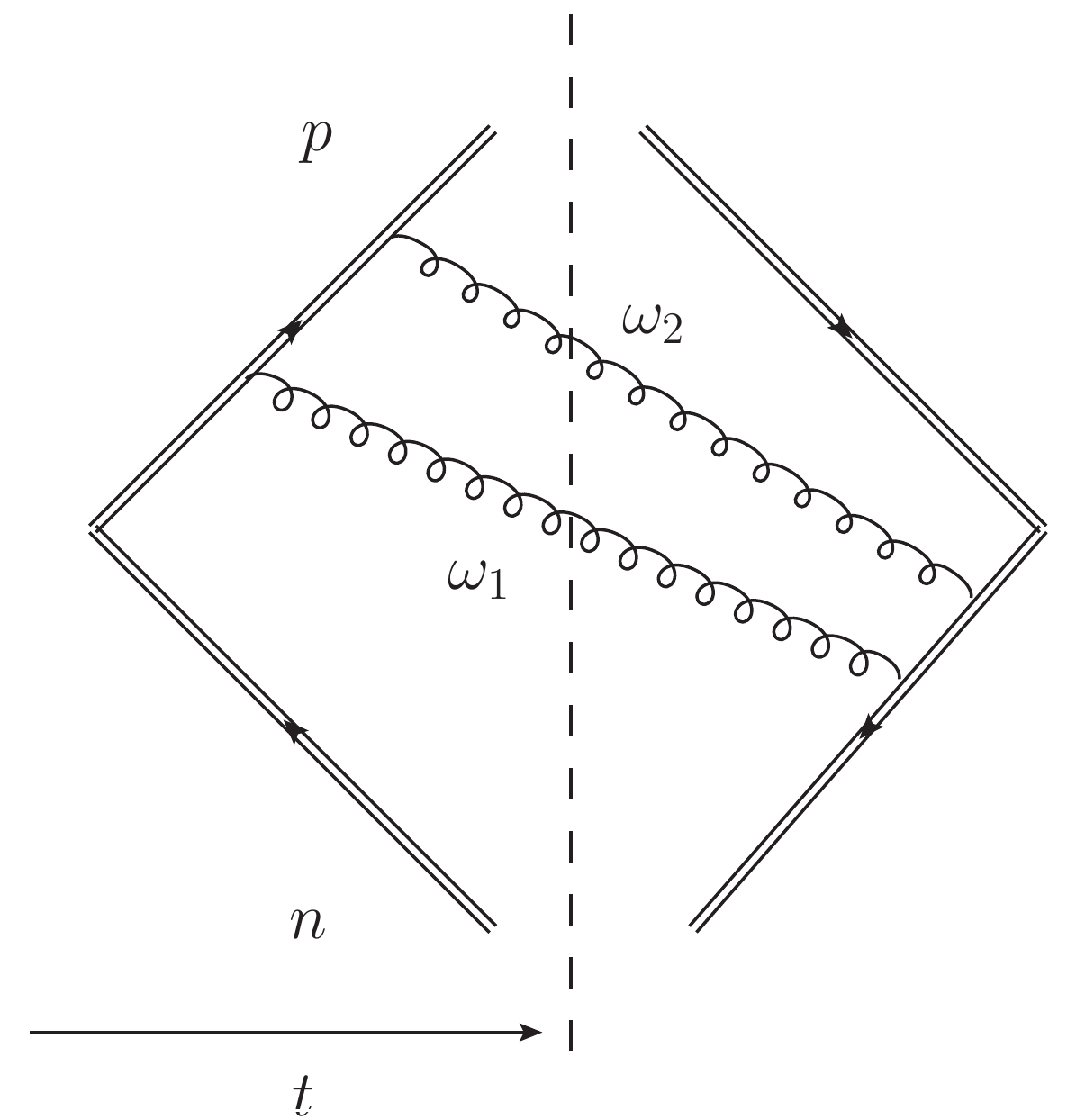}
 \caption{Another two gluon diagram which does not contribute to order $\alpha_s^2\ln^2 \frac{E}{E_0}$ for $\sigma_2$.}
  \label{fig:2gluonbms2}
\end{figure}

The BMS equation, describing the ``non-global'' logarithms in $e^+e^-$ annihilation process, is based on the universal feature of the underling branching process,
 where a large number of soft gluons is  released as  asymptotic final states, with  infrared divergent contributions to the total cross-section. In the eikonal approximation, which is sufficient for our purpose, the energetic quark-antiquark pairs are represented by Wilson-line cusp consisting of light-like gauge links with direction 4-vectors $p=(1,\vec{p})$ and $n=(1,\vec{n})$
\begin{align}
{\cal W}_{np}={\cal T} \bigg({\cal P}e^{ig\int_{0}^{\infty} ds p\cdot A(ps)}{\cal P}e^{ig\int_{\infty}^{0}ds n\cdot A(ns)}\bigg) \ .
\end{align}
The  amplitudes consisting of  $n$ soft gluons with momenta $k_1$,...$k_n$  emitted from the Wilson-line cusp ${\cal W}_{np}$ as final states, read
\begin{align}
\langle k_1,..k_n|{\cal W}_{np}|\Omega\rangle=iM_{n}(k_1,k_2,...k_n) \ .
\end{align}
The $n$-gluon contribution to the total-cross section,
\begin{align}
\sigma_n=\frac{1}{n!}\int d\Gamma_1d\Gamma_2...d\Gamma_n |M_{n}(k_1,k_2,...k_n)|^2 \ ,
\end{align}
normalizes to $1$, thanks to unitarity
\begin{align}
\sum_{n=0}^{\infty}\sigma_n=1 \ .
\end{align}
However, there are logarithmic IR divergences in $\sigma_n$,  represented as logarithms formed between the UV cutoff $E$ and the IR cutoff $E_0\gg \Lambda_{\rm QCD}$. We are interested in the ``most singular'' part of $\sigma_n$, namely, the part that a logarithm $\ln \frac{E}{E_0}$ always come with $\alpha_s$. It turns out that this part of the $\sigma_n$ can be effectively generated through a simple Markov process. More precisely
\begin{enumerate}
  \item In time-order perturbation theory, the emitted soft gluons are strongly ordered in energies, $E\gg \omega_1\gg\omega_2\gg...\omega_n\gg E_0$.
  \item Softer gluons are emitted later in time, harder gluons are emitted earlier in time.
  \item Each time a softer gluon with momentum $k$, is emitted from a harder gluon with momentum $p$, a factor $\frac{gp^{\mu}}{p\cdot k}$ for the emission kernel follows, in the eikonal approximation.
\end{enumerate}
It is easy to check that for emission processes violating one or more of the above properties, such as softer gluon emitted first, the contribution to $\sigma_n$ will be less singular. For an illustration of the energy and time ordering, see Figures.~\ref{fig:2gluonbms1} and~\ref{fig:2gluonbms2}.

We first consider the two-gluon diagram in Fig.~\ref{fig:2gluonbms1}.  In covariant perturbation theory,  the diagram is proportional to
\begin{align}
&\text{Fig.~\ref{fig:2gluonbms1}}=-\frac{1}{2}C_AC_F\int \frac{d^2\Omega_1 d^2\Omega_2}{(2\pi)^4}\int \frac{\omega_1^2\omega_2^2d\omega_1d\omega_2}{4\omega_1\omega_2(2\pi)^2}\frac{g^4}{(k_1+k_2)\cdot p k_2\cdot p (k_1+k_2)\cdot n}\nonumber \\
& \times \frac{-(2k_1+k_2)^{\mu}n^{\nu}+(2k_2+k_1)^{\nu}n^{\mu}+(k_1-k_2)\cdot n g^{\mu\nu}}{(k_1+k_2)^2}p_{\mu}p_{\nu} \ ,
\end{align}
where the color factor comes from $if^{abc}t^at^bt^c=\frac{i}{2}f^{abc}[t^a,t^b]t^c=-\frac{C_AC_F}{2}$. Now consider the region $|\vec{k}_1| \gg |\vec{k}_2|$, in this case the triple gluon vertex simplifies to $-2k_1^{\mu}n^{\nu}+k_1^{\nu}n^{\mu}+k_1\cdot n g^{\mu \nu}$. The last term vanishes due to $p^2=0$.
The second term proportional to $k_1^{\nu}$ (longitudinal) will cancel with contributions from polarization diagrams, after using Ward-identities, leaving only the physically motivated eikonal current $\frac{2k_1^{\mu}n^{\nu}}{2k_1\cdot k_2}$. As a result, the contribution in this region in the $C_F \sim \frac{N_c}{2}$ (large color) limit, is
\begin{align}\label{eq:2gluonfa}
&\text{Fig.~\ref{fig:2gluonbms1}}|_{\omega_1 \gg \omega_2}\nonumber \\
&=(\frac{\alpha_s N_c}{2\pi})\int \frac{d\omega_1}{\omega_1} \int \frac{d\Omega_1}{4\pi}\frac{p\cdot n}{\hat k_1 \cdot p \hat k_1\cdot n}(\frac{\alpha_s N_c}{2\pi})\int \frac{d\omega_2}{\omega_2}\int\frac{d\Omega_2}{4\pi}\frac{\hat k_1 \cdot p}{\hat k_2 \cdot \hat k_1 \hat k_2\cdot p} \ ,
\end{align}
with the desired factorized form, producing the leading logarithm. On the other hand, in the region $\omega_2 \ll \omega_1$, it is clear that the energy integral is
\begin{align}
\text{Fig.~\ref{fig:2gluonbms1}}|_{\omega_2 \gg \omega_1} \propto \int \frac{\omega_2^2d\omega_2}{2\omega_2\times\omega_2^4}\times \omega_2\int \frac{\omega_1^2d\omega_1}{2\omega_1\times \omega_1}=\int \frac{d\omega_2}{4\omega_2^2}\int d\omega_1=\int \frac{d\omega_2}{4\omega_2} \ ,
\end{align}
which only leads to a single logarithm. It is easy to check that in the time-ordered perturbation theory, the energy denominators are in one to one correspondence with  the eikonal propagators. For example in the region $\omega_1 \gg \omega_2$,  we have for the time order shown in Fig.~\ref{fig:2gluonbms1}
\begin{align}
&\frac{1}{\omega_1+\omega_2-\vec{p}\cdot(\vec{k_1}+\vec{k_2})}\frac{1}{\omega_2-\vec{p}\cdot \vec{k_2}}\frac{1}{\omega_1+\omega_2-|\vec{k_1}+\vec{k_2}|}\frac{1}{\omega_1+\omega_2-\vec{n}\cdot(\vec{k_1}+\vec{k_2})}
\nonumber \\ &\sim \frac{1}{k_1\cdot p}\frac{1}{k_2\cdot p}\frac{1}{k_2\cdot \hat k_1}\frac{1}{k_1\cdot n} \ .
\end{align}
The  contribution for another time ordering with backward moving gluon on the conjugate amplitude side,  is suppressed by more $\frac{1}{\omega_1}$ and is therefore less singular. Similarly, for Fig.~\ref{fig:2gluonbms2} one can show that in either $\omega_1 \gg \omega_2$ or $\omega_2 \gg \omega_1$ region, there will be no leading logarithm
\begin{align}
\text{Fig.~\ref{fig:2gluonbms2}}|_{\omega_1 \gg \omega_2} \propto \int \frac{\omega_1d\omega_1}{\omega_1^3}\int \frac{\omega_2 d\omega_2}{\omega_2}\propto \ln \frac{E}{E_0} \ , \\
\text{Fig.~\ref{fig:2gluonbms2}}|_{\omega_2 \gg \omega_1} \propto \int \frac{\omega_2d\omega_1}{\omega_2^3}\int \frac{\omega_1d\omega_1}{\omega_1}\propto \ln \frac{E}{E_0} \ .
\end{align}
In the first case the time-energy ordering is violated in the conjugate amplitude side, while in the second case the energy-time ordering is violated in the amplitude side. It is clear that the rule that softer gluon emitted later carries to higher orders. The  contribution for any leading $n$-gluon diagram will have a factorized form as in Eq.~(\ref{eq:2gluonfa}).

\begin{figure}[!htb]
\includegraphics[height=3.5cm]{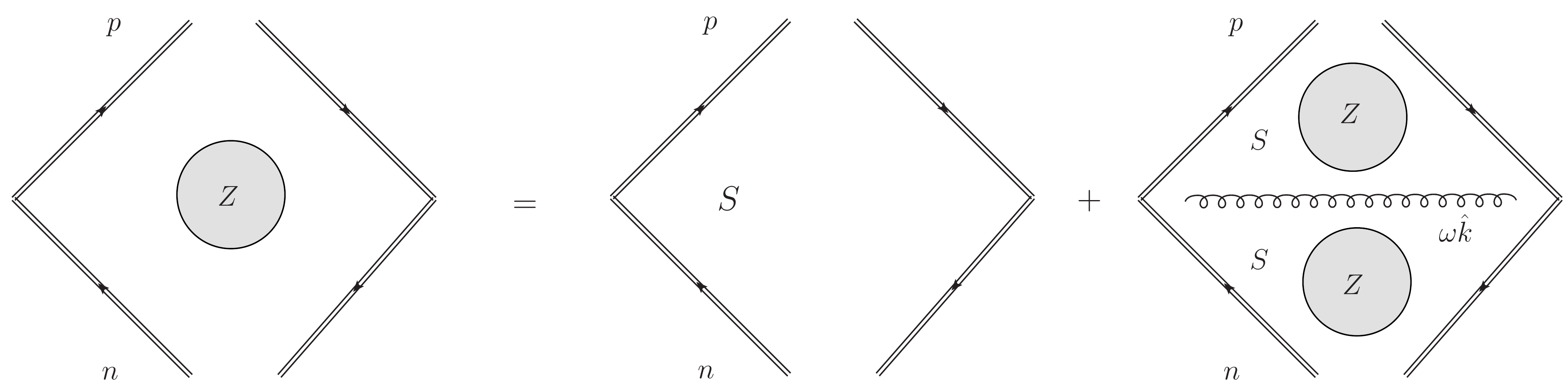}
 \caption{Illustration  of the recursive relation given in (\ref{eq:bms}). $S$ denotes the virtual Sudakov contribution. The first term after the equal sign corresponds to the pure virtual contribution to the generating function $Z$. The second term corresponds to the contribution with at leats one real gluon, the hardest one with momenta $k=\omega\hat k$ is emitted first and splits the original ``dipole'' into two.}
  \label{fig:bmsz}
\end{figure}

As a result, the emission depends only on the color-charges that are already present in the final state and their energy scales, but not on the history of how they are emitted. In large $N_c$,  the color charges effectively split the original Wilson-line cusp into many ``dipoles'', with the subsequent emissions to different dipoles being completely independent. To keep track of the above branching process,
we follow Mueller$^\prime$s reasoning for the wavefunction evolution, and introduce the generating functional
\bea
{\cal Z}\bigg(\frac{E}{E_0};n,p;\lambda\bigg)=\sum_{n=0}^{\infty}\lambda^n \sigma_n \ ,
\eea
for the n-cross-sections, which is readily seen to obey the integral equation
\bea
\label{eq:bms}
&&{\cal Z} \bigg(\frac{E}{E_0};n,p;\lambda \bigg)=e^{-\frac{\bar \alpha_s}{4\pi} \ln \frac{E}{E_0}\int d\Omega_k \frac{\hat p\cdot \hat n}{\hat k\cdot \hat p\hat k \cdot \hat n}}\nonumber \\
&&+\frac{\bar \alpha_s}{4\pi} \lambda \int_{E_0}^{E} \frac{d\omega}{\omega} e^{-\frac{\bar \alpha_s}{4\pi}\ln \frac{E}{\omega} \int d\Omega_k \frac{\hat p\cdot \hat n}{\hat k\cdot \hat p\hat k \cdot \hat n}}\int d\Omega_k \frac{\hat p \cdot \hat n}{\hat k\cdot \hat p\hat k \cdot \hat n}{\cal Z} \bigg(\frac{\omega}{E_0};n,\hat k;\lambda \bigg){\cal Z} \bigg(\frac{\omega}{E_0};p,\hat k;\lambda \bigg) \ ,
\eea
where $\bar \alpha_s= \frac{\alpha_sN_c}{\pi}$.
The first term is the Sudakov contribution where all soft gluons are virtual, and the second term is the contribution where at least one soft gluon is real (we pick the hardest gluon with momentum $k=\omega \hat k$, which splits the Wilson-line cusp into two dipoles at energy scale $\omega$, with or without real contributions). The pre-factor is $\frac{\alpha_sN_c}{\pi}$ instead of $\frac{\alpha_sN_c}{2\pi}$ as in Eq.~(\ref{eq:2gluonfa}),  because for each real gluon emission there are two contractions, whereas Fig.~\ref{fig:2gluonbms1} shows only one contraction for each of the two gluons.  See Fig.~\ref{fig:bmsz} for an illustration  of the recursive relation (\ref{eq:bms}). It is easy to check that for $\lambda=1$, one simply has $Z=1$, the required normalization property. The above is nothing but the integral form of the BMS equation~\cite{Banfi:2002hw,Marchesini:2003nh}. To cast it as a differential equation, one  takes the derivative with respect to $\ln \frac{E}{E_0}$
\begin{align}
&\frac{d}{d\ln \frac{E}{E_0}}{\cal Z}(\frac{E}{E_0};n,p;\lambda)\nonumber \\
&=\frac{\bar \alpha_s}{4\pi} \int d\Omega_k \frac{\hat p\cdot \hat n}{\hat k\cdot \hat p\hat k \cdot \hat n} \bigg(-{\cal Z}(\frac{E}{E_0};n,p;\lambda)+\lambda {\cal Z}(\frac{E}{E_0};n,\hat k;\lambda){\cal Z}(\frac{E}{E_0};p,\hat k;\lambda)\bigg) \ ,
\end{align}
which  is the standard form of the BMS equation in~\cite{Banfi:2002hw,Marchesini:2003nh}.

After introducing the BMS equation, we would like to provide a few comments from a field theoretical perspective. The ``power-expansion'' in asymptotically free quantum field theories is distinct from that for super-renormalizable or conformal field theories,  by additional logarithmic contributions that couple UV and IR scales\footnote{The standard operator product expansion implies that these logarithms are controlled asymptotically by perturbation theory, as they couple  both to IR and UV regimes.}.
In particular, the squared amplitudes for ``asymptotic-gluons'' integrated over their pertinent phase
space, suffers from logarithmic IR divergences with involved non-linear pattern as in~(\ref{eq:bms}). When these partial contributions are used as building blocks for other quantities, sometimes miracle cancellation occurs, and the logarithms in the final result follows a much simpler linear pattern. This normally happens when the IR scale $E_0$ is introduced through global conserved quantities,  such as the transverse momentum $Q_{\perp}$. However, in many other cases, the non-linear pattern of logarithms survives. This is the case for the  original ``BMS-non-global logarithm'' measuring the probability of energy flow less then $E_0$ outside a certain jet cone region $\Omega_{\rm in}$~\cite{Banfi:2002hw}, which is  not a conserved quantity. Another example is the total ``transverse energy'' $E_T=\sum_i \sqrt{\vec{k}^2_{i\perp}+m_i^2}$, for which factorization is known to break.

\subsection{Connecting BMS and Mueller's dipole}
The generating functional approach to the BMS equation makes the connection to small-$x$ physics very transparent. In fact, as shown by Mueller~\cite{Mueller:1993rr}, small-$x$ evolution equations such as BFKL, BK,  are all based on the distribution of virtual soft gluons in the light-front wave functions (LFWF). In the famous dipole picture, based on light-front perturbation theory, it is easy to show that the leading rapidity logarithms in the norm-squares of the dipole's LFWF can be effectively generated through a very similar Markov process, where the soft gluons emitted are strongly ordered in rapidity $x_1\gg x_2....\gg x_n$, and gluons with larger $x$ are emitted first. Based on this, one can write out the evolution equation for the generating function of the dipole distribution as in Sec.~\ref{SEC_DLA}.

Clearly, (\ref{eq:bms}) and (\ref{eq:evoZ}) are very similar, in the sense that they are both generating function for distribution of soft gluons in ``final states'', and both originates from a Markov branching process with the gluons  strongly ordered in the evolution variables.  But, there are differences.  The Mueller's evolution equation is for a wave function, and the evolution is towards the direction with more and more ``virtuality'' ($\frac{k_{1\perp}^2}{2x_1}\ll ...\ll \frac{k_{n\perp}^2}{2x_n}$), while in BMS the evolution is towards more and more ``reality''. This suggests that the mapping between the BMS and Mueller$^\prime$s evolutions, flips the energy scale. This is only possible in the  conformal limit.

Indeed, one can perform a conformal transformation to map the Mueller's dipole construct, to exactly the Wilson-line cusp~\cite{Cornalba:2007fs,Vladimirov:2016dll}. The transformation reads
\begin{align}\label{eq:conformal}
(x^+,x^-,\vec{x}_\perp)\ \ {\cal C}: \rightarrow  \ \ (x^+-\frac{x_\perp^2}{2x^-},-\frac{1}{2x^-},\frac{\vec{x}_\perp}{\sqrt{2}x^-}) \ .
\end{align}
It is easy to check that (\ref{eq:conformal})  maps a Wilson-line cusp ${\cal W}_{np}$ pointing to positive infinity, onto a dipole propagating along  the LF time $x^+$ from $x^+=-\infty$ to $x^+=0$
\begin{align}
{\rm P}\exp \bigg[ig\int_{0}^{\infty} ds v\cdot A(sv)\bigg] \ \ {\cal C}: \rightarrow  \ \ {\rm P}\exp \bigg[ig\int_{-\infty}^{0} dx^+ A^-(x^+,0,\frac{\vec{v}_\perp}{2v^-})\bigg] \ .
\end{align}
The asymptotic soft gluons emitted into the scattering states at $t=\infty$,  become the virtual gluons present in the wave function at zero light-front time $x^+$! In this sense, the mapping is a ``virtual-real'' duality, in addition to the standard interpretation that it maps rapidity divergences to UV divergences~\cite{Vladimirov:2016dll}\footnote{In non-conformal theory, the exact mapping breaks at two-loop already~\cite{Caron-Huot:2015bja}. But for the virtual part there is a way to generalize the mapping to all orders~\cite{Vladimirov:2016dll}.}. Moreover, if one parameterizes the direction vector as $v=v^0(1,\vec{n})$, then the conformal transformation reduces to the standard stereographic projection, that maps $\vec{n}\in {\rm S}^2$ onto $\vec{b} \in {\rm R}^2$. See Table.~\ref{tab:comparasion} for a comparison between the BMS and dipole evolution.

\begin{table*}
\caption{\label{tab:comparasion} Mueller hierarchy and BMS hierarchy}
\begin{tabular}{ c|c|c}
\hline
& dipole& Cusp \\
\hline\hline
distribution in& {\it virtual} gluon in wave function & {\it real} gluon in asymptotic state\\
\hline
large $N_c$& yes&yes  \\
\hline
evolution in & rapidity divergence&soft divergence \\
\hline
kernel&$\frac{b^2_{10}}{b^2_{12}b^2_{20}}$& $\frac{n\cdot p}{\hat k\cdot n \hat k\cdot p}$ \\
\hline
virtual part& TMD soft factor& Sudakov form factor  \\
\hline
time ordering& in LF time $x^+$&in CM time $t$  \\
\hline
momentum ordering&decreasing $k^+$& decreasing energy $\omega$ \\
\hline
virtuality ordering& increasing&decreasing   \\
\hline
Markov property&yes&yes  \\
\hline
DLA & $b_{10}\gg b_{12} \gg ...$& $\theta_{01}\gg \theta_{12}\gg....$    \\
\hline
\end{tabular}
\end{table*}

\begin{figure}[!htb]
\includegraphics[height=9cm]{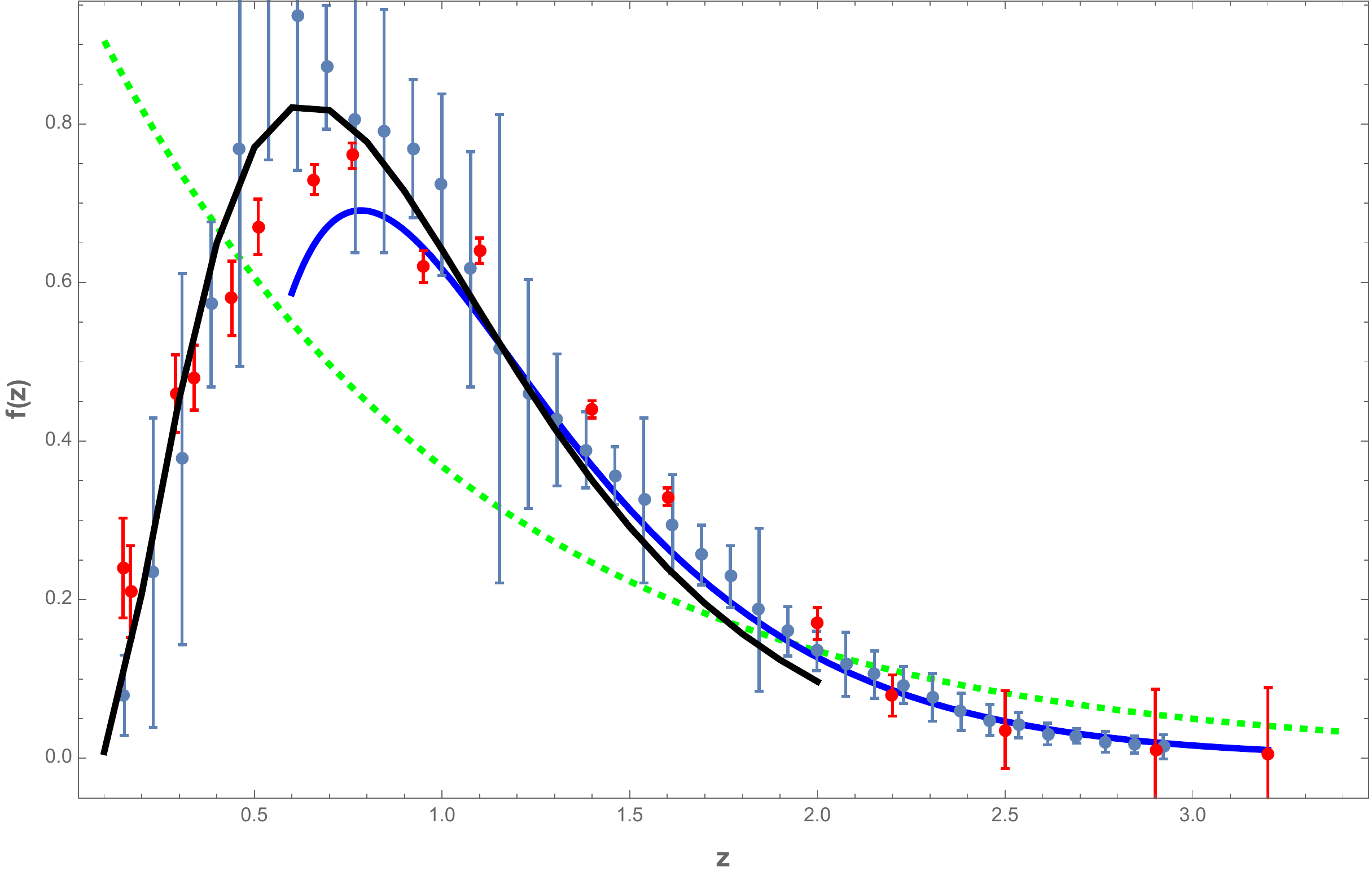}
 \caption{The exact scaling   (\ref{FZALL})  of the KNO particle multiplicity (solid-black),  its exact asymptotic (\ref{eq:fsamylarge})
 (solid-blue) and the diffusive KNO multiplicity (green-dashed),
 are compared to the measured particle multiplicities in ep DIS scattering by the H1 collaboration~\cite{H1:2020zpd} (red data points),
 and the charged particle multiplicities from hadronic Z-decays from $e^+e^-$ annihilation with $\sqrt{s}=91.2$ GeV, as measured by the ALEPH collaboration~\cite{ALEPH:1995qic} (grey data points).}
  \label{fig:EEFZ}
\end{figure}

\subsection{The universal DLA limit}
The BK/BMS equation, resums single logarithms in rapidity/energy. However, in both cases there are two types of divergences instead of one: in the Mueller's dipole there are UV divergences (in $p_n$) in large $k_\perp$ or small dipole region, while in the Wilson-line cusps there are ``rapidity-divergences'' caused by emissions collinear to the Wilson-lines. It is natural to consider the double logarithms (DL) in terms of both\footnote{In fact, it is known that the Sudakov form factor is dominated by double logarithm in the exponential.}. One can show that the leading double logarithms are generated from the same branching process, that in addition to the strong ordering in $k^+$/$\omega$, one imposes strong ordering in dipole sizes/emmiting angles as well. The strong ordering in virtuality is preserved by the DL limit.  Clearly, the two DL limits and the underlying size/angle orderings,  map onto each other under the conformal transformation (\ref{eq:conformal}). We  emphasize that the DL limit dominates the large $Q^2$ limit of the $e^+e^-$ multiplicity
\begin{align}
n_{e^+e^-}(Q^2)=\exp \bigg[2\bigg(\frac{N_c}{2\pi \beta_0}\ln \frac{Q^2}{\Lambda_{\rm QCD}^2} \ln \frac{\ln \frac{Q^2}{\Lambda_{\rm QCD}^2}}{\ln \frac{Q_0^2}{\Lambda_{\rm QCD}^2}}\bigg)^{\frac 12}\bigg] \ ,
\end{align}
while the BFKL limit~\cite{Marchesini:2003nh} only contributes to
$$n_{\rm BFKL} \sim e^{\frac{N_c\ln 2}{\pi \beta_0} \ln \frac{\ln \frac{Q^2}{\Lambda_{\rm QCD}^2}}{\ln \frac{Q_0^2}{\Lambda_{\rm QCD}^2}} } \ ,$$
smaller than the DLA result in the large $Q^2$ limit.

In Fig.~\ref{fig:EEFZ} we show the exact scaling (\ref{FZALL})  of the KNO particle multiplicity (solid-black),  versus
the measured multiplicities in ep DIS scattering by the H1 collaboration~\cite{H1:2020zpd} (red data points),
 and the charged particle multiplicities from hadronic Z-decays from $e^+e^-$ annihilation with $\sqrt{s}=91.2$ GeV,
 as measured by the ALEPH collaboration~\cite{ALEPH:1995qic} (grey data points). We have also shown
 the exact asymptotic (\ref{eq:fsamylarge})  (solid-blue) and the diffusive KNO multiplicity (green-dashed) for
 comparison.  The agreement is very good, for both
 data sets, underlying the universality of our results.

\section{Implication for the entanglement entropy}
\label{SEC_ENT}
As we have shown in~\cite{Liu:2022hto}, the reduced density matrix measuring quantum entanglement between fast and slow degrees of freedom in Mueller$^\prime$s dipole wavefunction, has the generic form
\begin{align}
\label{SCH}
\rho=\sum_{n} p_n\rho_n\ ,
\end{align}
where $p_n$ is the total probability of finding $n$-dipoles, and the $\rho_n$ is an effective reduced density matrix with $n$-soft gluon on the left and right. From (\ref{SCH}) the entanglement entropy for $\rho$ can be found to be
\begin{align}
S=-\sum_n p_n \ln p_n +\sum_{n} p_n s_n \ ,
\end{align}
where $s_n=-{\rm tr}\rho_n \ln \rho_n$ is the entanglement entropy of the reduced density matrix in the $n$-particle sector. Since the wave function peaks  at $n=\bar n$ for large $n$, it is natural to expect that the entanglement entropy $s_n$ is also peaked around $n=\bar n$,  and scales as $s(\frac{n}{\bar n})$. Under this assumption, the universal behavior follows
\begin{align}\label{EEKNO}
S=\ln \bar n+\int dz f(z)\bigg(-\ln f(z)+s(z)\bigg) \ .
\end{align}
In particular, in  the DLA regime, the use of the asymptotic form of (\ref{NRHO}) yields
\bea \label{eq:entropy}
S(\bar{n}(y,Q^2))\rightarrow  {\rm ln}(\bar{n}(y,Q^2))\equiv
2\bigg(\frac{N_c}{\pi \beta_0}y \ln \frac{\ln \frac{Q^2}{\Lambda_{\rm QCD}^2}}{\ln \frac{Q_0^2}{\Lambda_{\rm QCD}^2}}\bigg)^{\frac 12} \ .
\eea
with $\beta_0$ fixed in (\ref{AG}). The region of validity for the DLA implies
\begin{align}
\label{TWOLL}
y \sim \ln \frac{\ln \frac{Q^2}{\Lambda_{\rm QCD}^2}}{\ln \frac{Q_0^2}{\Lambda_{\rm QCD}^2}} \rightarrow \infty \ ,
\end{align}
which shows that the largest logarithm is  the double logarithm. More explicitly,
(\ref{TWOLL}) implies that $Q^2$ must be very large, putting the saturation regime out of reach in the DLA regime.
The entanglement entropy (\ref{eq:entropy}) is accessible to current and future DIS measurements.

We note that although both the DLA and diffusive regimes, support  KNO scaling, the corresponding scaling functions are very different.
In the diffusive limit, the distribution $p_n \rightarrow e^{-n/\bar n}$,  is almost identical to the thermal distribution for a quantum oscillator with $\bar n\sim \frac{T}{\omega}\gg 1$,  which suggests  maximal decoherence encoded in the large entanglement entropy $S\sim  y$. However, the multiplicity distribution in the DLA regime, is far from thermal, with a much smaller entanglement entropy $S\sim  \sqrt{y}$. Also, its KNO scaling function $f(z)$ carries still  more structure (peak at intermediate $z$).

Finally, due to the similarity in the branching process underlining the BMS evolution and Mueller's dipole, one can define a reduced density matrix that entangles soft gluons in the final state at different energy scales for the leading order BMS evolution. It satisfies a similar evolution equation that maps exactly to that for the Mueller's dipole as we have shown in~\cite{Liu:2022hto}. In the large $Q^2$ limit, it produces a large entanglement entropy, responsible for the large observed multiplicities in jets. In the DLA limit, the entanglement entropy in jet emissivities  is again given by a result similar to (\ref{eq:entropy}),
\begin{align}
\label{SCUSP}
S_{\rm cusp}(\bar{n}(Q^2))\rightarrow  {\rm ln}(\bar{n}(Q^2))\rightarrow
2\bigg(\frac{N_c}{2\pi \beta_0}\ln \frac{Q^2}{\Lambda_{\rm QCD}^2} \ln \frac{\ln \frac{Q^2}{\Lambda_{\rm QCD}^2}}{\ln \frac{Q_0^2}{\Lambda_{\rm QCD}^2}}\bigg)^{\frac 12} \ .
\end{align}
The rapidity gap between the quark-antiquark pair $y=\ln \frac{Q^2}{\Lambda_{\rm QCD}^2}$,  produces another logarithm in $Q^2$. The argument of the square root
in (\ref{SCUSP}) is  the Sudakov double logarithm $$|H_{\rm suda}(Q^2)|^2\sim e^{-\frac{N_c}{2\pi \beta_0}\ln \frac{Q^2}{\Lambda_{\rm QCD}^2}\ln\ln \frac{Q^2}{\Lambda_{\rm QCD}^2}} \ , $$ in the large $Q^2$ limit, namely
\bea
S_{\rm cusp}(\bar{n}(Q^2))\rightarrow  2\sqrt{2}\,\bigg({\rm ln}\bigg(\frac 1{|H_{\rm suda}(Q^2)|}\bigg)\bigg)^{\frac 12} \ .
\eea
The entanglement entropy (\ref{SCUSP}) is also accessible to currently measured emissivities from jets, say from $e^+e^-$ annihilation. As mentioned before, in case of $e^+e^-$ the DLA dominates the large $Q^2$ limits in comparison to the BFKL contribution.

\section{Conclusion and Outlook}
\label{SEC_CON}
We have presented an exact derivation of the leading moments, of the dipole emissivities from the dipole cascade originating
from Mueller$^\prime$s wave function evolution in $1+3$ dimensions, by re-summing the leading logarithms
in both large $Q^2$ and large rapidity $y={\rm ln}\frac 1x$, which we refer to as the DLA limit.
We have shown that the hierarchy of moments, allows the reconstruction of a continuous probability distribution $f(z)$ supported in $(0,\infty)$, which is nothing but the
KNO scaling function of the dipole multiplicity $p_n=\frac{1}{\bar n}f(\frac{n}{\bar n})$ with $z=\frac{n}{\bar n}$.  The behavior of $f(z)$ at large and small $z$ can be exactly determined, while the
result for all values of $z$ is only accessible   numerically.

In principle, the virtual dipoles in the LFWF still need to pass through the final state evolution stage, to become real asymptotic states. However, the final state evolution is
less rapidity-divergent, and we expect that the main features of the dipole multiplicity distribution to hold. This distribution  can be used as a probe for the final  hadron multiplicities, as suggested in~\cite{Gotsman:2020bjc,Kharzeev:2021yyf} (and references therein). Indeed, our parameter free result reproduces well the DIS multiplicities reported by the H1 collaboration at HERA,
in the highest $Q^2$ range, in particular, the observed KNO scaling function is in good agreement with the predicted dipole scaling function, including the overall shape, the location of the peak and the large $z=\frac n{\bar{n}}$ tail. Our results show that the currently
available DIS data at HERA are more amenable to the present DLA regime, than the diffusive or BFKL regime. The
gluon multiplicities both for large and small $z=\frac n{\bar{n}}$ in the DLA regime, should prove useful for more detailed comparisons with
present and future DIS data, at large $Q^2$ and small parton-$x$. We have provided a pedagogical introduction to the leading order BMS evolution equation and its relation to dipole evolution, from which the universality of the KNO scaling function is manifest.

The entanglement entropy in the DLA regime of DIS, is  found to asymptote  $S={\rm ln}\bar n$,  much like in the diffusive (BFKL) regime.
This is a general result of KNO scaling of the ensuing gluon multiplicities, satisfied by both regimes.
However, the growth  of the mean multiplicity $\bar n=xG(x,Q^2)$,
is slower with $\bar n\sim e^{\#\sqrt{\alpha_s\,y}}$
or $S\sim \sqrt{\alpha_s\,y}$ in the DLA regime,  and faster
with  $\bar n\sim e^{\#\alpha_s\,y}$ or $S\sim \alpha_s\,y$ in the BFKL regime.
We regard this as an indication of faster scrambling of quantum coherence in the diffusive regime
(smaller and smaller off-diagonal entries in the entangled density matrix).
We emphasize that the entanglement entropy for DIS and jets in the DLA regime, is directly accessible to
current and future measurements.

Needless to say, that the information encoded in the QCD multiplicities in the DLA regime,
is far more detailed than that captured by the  entanglement entropy. However, the latter
can be used as a  sharp  characterization of saturation,
where the  quantum cascade of dipoles  reaches a state of maximum decoherence. Recall that
a pure state with maximal coherence, carries zero entanglement entropy.
But what is the signature for this onset, and bound if any?

Saturation as a regime of maximal decoherence in the QCD cascade evolution of dipoles as gluons,
is likely to take place in the  diffusive rather than DLA regime,  where the entanglement entropy is substantially larger
with increasing rapidity. This is further supported by the observation that the diffusive multiplicities at weak coupling, are
very similar to the thermal distribution of a quantum oscillator. It is therefore not surprising, that
the same quantum entropy was noted in the dual string (a collection of quantum oscillators)
exchanged between highly boosted hadrons (with emergent Unruh temperature).
The dual string quantum entropy is extensive with the
rapidity, commensurate with the transverse size growth of the
boosted hadron as a stretched string, and saturates at one bit per string length squared~\cite{Liu:2018gae}.
These are the signature and bound, we are looking for, in characterizing the emitted hadronic multiplicities,
as measured in both DIS and hadron scattering with large rapidity gaps.
Remarkably, a similar signature  and bound  are observed in a classical black-hole, where
information  is maximally scrambled on its near horizon, and saturates to  the lowest bound of one bit per Planck length squared~\cite{Bekenstein:1973ur}.


\vskip 1cm

{\bf Acknowledgements}

We are grateful to Jacek Wosiek for bringing to our attention~\cite{Dokshitzer:1991wu} and to Yoshitaka Hatta  for informing us about~\cite{Weigert:2003mm,Hatta:2013iba}. This work is supported by the Office of Science, U.S. Department of Energy under Contract No. DE-FG-88ER40388 and by the Priority Research Areas SciMat and DigiWorld under program Excellence Initiative - Research University at the Jagiellonian University in Krak\'{o}w.

\appendix
\section{Multiplicity distributions in super-renormalizable theories}
In this Appendix we show in two examples, that the  multiplicity distribution in super-renormalizable theories
(real or virtual), is in general Gaussian-like in the scaling region. This behavior is similar  to the multiplicity
distribution observed for the onium$^\prime$s LFWF in $1+2$D QCD~\cite{Liu:2022hto}.
In other words,  KNO scaling holds only for critical theories
 with dimensionless couplings~\cite{Polyakov:1970lyy,Balog:1997xr}.
\\
\\
{\bf Ising model:}
\\
The first example is the multiplicity distribution in the massive continuum limit of the 2D Ising model at zero magnetic field~\cite{Wu:1975mw,McCoy:1977er}, where the mass of the free-fermion equals to $m$. We consider the spin-spin correlator $G(r)=\langle \Omega|\sigma(r)\sigma(0)|\Omega\rangle$ in its ``form-factor expansion''~\cite{Wu:1975mw,McCoy:1977er,Berg:1978sw}
\begin{align}
G(r)=\sum_{n}\int \prod_i\frac{dp_i}{4\pi E_i}|\langle \Omega|\sigma(0)|p_1..p_n\rangle|^2e^{-r\sum_{i=1}^n E(p_i)} \ .
\end{align}
The relative contribution from the $n$-particle sector can be regarded as a multiplicity distribution
\begin{align}
\sigma_n(r)=\frac{1}{G(r)}\int \prod_i\frac{dp_i}{4\pi E_i}|\langle \Omega|\sigma(0)|p_1..p_n\rangle|^2e^{-r\sum_{i=1}^n E(p_i)} \ ,
\end{align}
with  the normalization  $\sum_n \sigma_n(r)=1$. To analyse this distribution at small $mr \ll 1$ or large energy $Q\gg m$, we consider the   ``lambda-extension''~\cite{McCoy:1976cd}
\begin{align}
G(r;\lambda)=\sum_{n}\lambda^n \int \prod_i\frac{dp_i}{4\pi E_i}|\langle \Omega|\sigma(0)|p_1..p_n\rangle|^2e^{-r\sum_{i=1}^n E(p_i)} \ ,
\end{align}
in terms of which the generating function for $\sigma_n$,  is
\begin{align}
\label{ZLAMBDA}
Z(\lambda,r)=\frac{G(r;\lambda)}{G(r)} \ .
\end{align}
For $\lambda\leq 1$, $G(r;\lambda)$ has representation in terms of special solutions to Painlevé equations~\cite{Wu:1975mw,McCoy:1976cd}, with a small-$r$ asymptotic~\cite{McCoy:1976cd,Tracy:1990tn}
\begin{align}
G(r;\lambda)|_{r\rightarrow 0}=\tau_0(\sigma)(mr)^{-\frac{\sigma}{2}(1-\frac{\sigma}{2})} \ ,
\end{align}
where we have defined
\bea
&&\sigma \equiv \sigma (\lambda)=\frac{2}{\pi} {\rm Arcsin}(\lambda) \ , \\
&&\tau_0(\sigma)=Ae^{-3s^2\ln 2}\Gamma_2(1+s)\Gamma_2(1-s) \ \ ,  \ s=\frac{1-\sigma}{2} \ .
\eea
Here $\Gamma_2(z)$ is the Barnes-$G$ function, and the value of the $\lambda$-independent constant $A$ is not important.  The mean multiplicity in the small-$r$ limit  follows as \begin{align}
\langle n\rangle=\frac{d}{d\lambda}Z(\lambda,r)|_{\lambda=1} \rightarrow \frac{2\ln \frac{8e^{\gamma_{\rm E}+1}}{mr}}{\pi^2}+{\cal O}\left(mr \ln^2 mr \right)\sim \frac{2}{\pi^2} \ln \frac{1}{mr}  \ ,
\end{align}
which is seen to grow  logarithmically. On the other hand, the mean square deviation of the distribution reads
\begin{align}
\sigma^2=\langle n^2\rangle-\langle n\rangle^2=\frac{d^2}{d^2\lambda}Z(\lambda,r)|_{\lambda=1} +\langle n\rangle -\langle n\rangle^2 \rightarrow \frac{4 }{3\pi^2}\ln \frac{1}{mr} \sim \frac{2}{3}\langle n \rangle\ .
\end{align}
Therefore, the width is of order $\sqrt{n}$, typical for a Poisson-like distribution.  More specifically, we have
\begin{align}
\lim_{r\rightarrow 0}Z\left(1-\frac{t}{\sigma},r\right)\left (1-\frac{t}{\sigma} \right)^{-\langle n \rangle}= e^{\frac{t^2}{2}} \ ,
\end{align}
which implies
\begin{align}
\sigma_n \rightarrow \frac{1}{\sigma}e^{-\frac{(n-\bar n)^2}{2\sigma^2}} \ .
\end{align}
The multiplicities fall into the  Poisson universality class.
\\
\\
{\bf phi-4:}
\\
The second example is the ground state wave function in terms of a free Fock-basis in a super-renormalizable theory. As an example, consider a 2D free massive scalar field $\phi$, supplemented by the quartic  interaction term $H_{\rm I}=g\int_{-\frac{L}{2}}^{\frac{L}{2}} :\phi^4(x): dx$. In the Fock-basis of the free-theory, the ground state wave function of the interacting theory can be expanded as
\begin{align}
|\Omega(L)\rangle=\sum_{n}p_n(L) |\Psi_n\rangle \ , \ \langle \Psi_n|\Psi_n\rangle=1 \ ,
\end{align}
where $p_n(L)$ is the probability of finding $n$-particles of the free-theory. As $L \rightarrow \infty$, the average particle number goes  to infinity, and we would like to find out the distribution of $p_n$ in this limit. For that, we introduce the generating function
\begin{align}
Z(\lambda,L)=\sum_{n}\lambda^n p_n(L) \ .
\end{align}
In  the large $L$ limit,  as  each connected diagram contributes  a single factor of $L$, and the disconnected diagrams factorize (after summing over all time orderings), we have
\begin{align}
Z(\lambda,L)=e^{-LF(1,g)}e^{L F(\lambda,g)} \ ,
\end{align}
where $L F(1,g)$ is  the ``field renormalization factor'' for the vacuum. Here,  $F(\lambda,g)$ is the sum over all the connected ``real-contributions'' weighted over the number of particles in the ``cut''. Clearly, when $\lambda=1$, $Z=1$. For large $L$, we have
\bea
&&\bar n=L\frac{dF(\lambda,g)}{d\lambda}|_{\lambda=1} \ , \\
&&\sigma^2=L\frac{d^2F(\lambda,g)}{d^2\lambda}|_{\lambda=1}+L\frac{dF(\lambda,g)}{d\lambda}|_{\lambda=1}  \ .
\eea
With this in mind, we can expand the generating function  around $\lambda=1$, with the result
\begin{align}
\lim_{L\rightarrow \infty}Z\left(1-\frac{t}{\sigma},L\right)\left (1-\frac{t}{\sigma} \right)^{-\langle n \rangle}= e^{\frac{t^2}{2}} \ .
\end{align}
Again, this implies a Gaussian distribution for  the limiting $p_n$.

\bibliography{ENT}

\end{document}